\documentclass[%
 aip,
 amsmath,amssymb,
 reprint,
]{revtex4-2}

\usepackage{graphicx}
\usepackage{dcolumn}
\usepackage{bm}

\usepackage[utf8]{inputenc}
\usepackage[T1]{fontenc}
\usepackage{mathptmx}
\usepackage{etoolbox}

\makeatletter
\def\@email#1#2{%
 \endgroup
 \patchcmd{\titleblock@produce}
  {\frontmatter@RRAPformat}
  {\frontmatter@RRAPformat{\produce@RRAP{*#1\href{mailto:#2}{#2}}}\frontmatter@RRAPformat}
  {}{}
}%
\makeatother

\DeclareFontFamily{U}{mathb}{\hyphenchar\font45}
\DeclareFontShape{U}{mathb}{m}{n}{
      <5> <6> <7> <8> <9> <10> gen * mathb
      <10.95> mathb10 <12> <14.4> <17.28> <20.74> <24.88> mathb12
      }{}
\DeclareSymbolFont{mathb}{U}{mathb}{m}{n}
\DeclareFontSubstitution{U}{mathb}{m}{n}
\DeclareMathSymbol{\circlearrowleft}       {3}{mathb}{"F6}
\DeclareMathSymbol{\circlearrowright}      {3}{mathb}{"F7}
\DeclareMathSymbol{\curvearrowleftright}   {3}{mathb}{"F2}

\newcommand\inter{\curvearrowleftright}

\newcommand{\intra}{{\mathchoice
    {
        \hspace{-0.7ex}
        {\mathrel{{\ooalign{\hss\raisebox{1.2ex}{
            \rotatebox{180}{$\circlearrowright$} 
            }\hss\cr\raisebox{1.2ex}{
            \rotatebox{180}{$\circlearrowleft$} 
            }}}}
        \hspace{-0.7ex}
        }
    }
    {
        \hspace{-0.7ex}
        {\mathrel{{\ooalign{\hss\raisebox{1.2ex}{
            \rotatebox{180}{$\circlearrowright$} 
            }\hss\cr\raisebox{1.2ex}{
            \rotatebox{180}{$\circlearrowleft$} 
            }}}}
        \hspace{-0.7ex}
        }
    }
    {
        \hspace{-1.5ex}
        {\mathrel{{\ooalign{\hss\raisebox{0.9ex}{
            \rotatebox{180}{
                \scalebox{0.7}{
                    $\circlearrowright$
                }
            } 
            }\hss\cr\raisebox{0.9ex}{
            \rotatebox{180}{
                \scalebox{0.7}{
                    $\circlearrowleft$
                }
            } 
            }}}}
        \hspace{-1.5ex}
        }
    }
    {
        \hspace{-1.3ex}
        {\mathrel{{\ooalign{\hss\raisebox{0.7ex}{
            \rotatebox{180}{
                \scalebox{0.6}{
                    $\circlearrowright$
                }
            } 
            }\hss\cr\raisebox{0.7ex}{
            \rotatebox{180}{
                \scalebox{0.6}{
                    $\circlearrowleft$
                }
            } 
            }}}}
        \hspace{-1.3ex}
        }
    }
}
}

\usepackage{hyperref}
\usepackage{color} 
\usepackage{comment}

\newcommand{\nn}{\nonumber \\}

\newcommand{\Wnhse} { W^{\rm nhse} }

\begin{document}

\preprint{AIP/123-QED}

\title[ ]{Higher-winding phases in one-dimensional non-Hermitian topological superconductors }

\author{Yung-Yeh Chang}
\affiliation{Institute of Physics, Academia Sinica, Taipei 115201, Taiwan}

\author{Xiang-Yu Li}
\affiliation{Institute of Physics, Academia Sinica, Taipei 115201, Taiwan}
\affiliation{Department of Physics, National Taiwan University, Taipei 10617, Taiwan}

\author{Ken Shiozaki}
\affiliation{Yukawa Institute for Theoretical Physics, Kyoto University, Kyoto 606-8502, Japan
}%

\author{Chen-Hsuan Hsu} 
\affiliation{Institute of Physics, Academia Sinica, Taipei 115201, Taiwan}
\affiliation{Physics Division, National Center for Theoretical Sciences, Taipei 106319, Taiwan}

\date{\today}

\begin{abstract}

Non-Hermitian topological superconductors provide a setting in which point-gap topology, non-Hermitian skin effects, and Majorana zero modes are strongly intertwined. In this work, we adopt a coefficient-based approach for computing winding numbers and deriving analytical expressions for phase boundaries in one-dimensional non-Hermitian topological superconductors characterized by point-gap topology with $\mathbb{Z}$ invariants. We apply this approach to two non-Hermitian topological superconducting lattice models, with and without sublattice degrees of freedom, including longer-range hoppings, thereby accessing a much broader parameter space. These extensions generate higher-order polynomials and support phases with higher winding numbers, reflecting the underlying $\mathbb{Z}$ topology. We further clarify how a weak perturbation suppresses the  non-Hermitian skin effect while preserving the sublattice-symmetry-protected invariant associated with Majorana zero modes. The predicted winding numbers are verified by open-boundary spectra, where one or multiple pairs of zero-energy boundary modes appear consistently with the bulk invariant. We also examine the stability of these modes against onsite disorder by examining the zero-mode energy, the bulk gap, and
the inverse participation ratio. Our results provide a systematic and efficient route to constructing topological phase diagrams for higher-winding non-Hermitian topological superconductors.

\end{abstract}

\maketitle

\section{Introduction}
\label{sec:intro}

Topological phases are characterized by topological invariants, and their phase transitions are signaled by changes in these invariants.~\cite{TKNN:1982,Hasan:2010,Qi:2011} Non-Hermitian systems~\cite{Hatano:1996,Hatano:1997,Bender:2003,Bender:2007,Brody:2014}   have recently emerged as a natural setting for extending this paradigm beyond the conventional Hermitian framework. The classification of non-Hermitian topological phases~\cite{Gong:2018,Kawabata:2019,Bergholtz:2021,Okuma:2023,shiozaki:2026} 
generalizes the Hermitian classification~\cite{Ryu:2010,Hasan:2010,Qi:2011} by incorporating symmetry structures unique to non-Hermitian Hamiltonians. This broader framework reveals phenomena without direct Hermitian counterparts. A prominent example is boundary-condition sensitivity: spectra and eigenstates under open boundary conditions (OBC) can differ drastically from those under periodic boundary conditions (PBC), leading to the non-Hermitian skin effect (NHSE), in which an extensive number of eigenstates become localized near the boundary.~\cite{Yao:2018,Yokomizo:2019,Lee:2019,Borgnia:2020,Kawabata:2020,Okugawa:2020,Okuma:2020,Yoshida:2020,Okugawa:2021,Zhang:2021,Schindler:2023,Hamanaka:2023,Manna:2023,Nakamura:2023,Yoshida:2024,Hamanaka:2024,Hetenyi:2025a,Wang:2025,Saito:2026,Phan:2026,Pu:NHISE-PRL-2026} 
This boundary accumulation is closely tied to point-gap topology, which provides a topological characterization of the NHSE.~\cite{Yao:2018,Borgnia:2020,Okuma:2020}

Recent work has also explored one-dimensional (1D) non-Hermitian topological superconducting (NHTSC) models,~\cite{Kawabata:2018b,Avila:2019,Okuma:2019,Zhao:PRB2021_nhKitaevChain,Arouca:2023,Mondal:2024,Chang:2026} 
where Majorana zero modes (MZMs) can appear under certain conditions. When the NHSE is suppressed,~\cite{Okuma:2019,Chang:2026} the winding number and gap-closing conditions derived from the PBC Hamiltonian correctly predict the number of MZMs under OBC and the corresponding phase boundaries, respectively. This provides a useful setting in which PBC topological invariants can be directly compared with OBC boundary modes and used to  establish reliable phase diagrams.

While the symmetry classification of these models permits a $\mathbb{Z}$ invariant, most regimes explored so far involve nontrivial phases with only a single pair of MZMs. Higher-winding phases remain less explored. A natural route toward such phases is to include longer-range hopping or pairing terms,~\cite{Rafi-Ul-Islam:2024} as in Hermitian topological superconductors.~\cite{Niu:2012,Russo:2013} However, these terms introduce higher harmonics in the PBC Hamiltonian and lead to higher-order polynomials in the expression of the winding number, making it difficult to extract winding numbers and phase boundaries analytically from the PBC spectrum. 

The aim of this work is to develop a coefficient-based approach for analyzing higher-winding phases in 1D NHTSCs. Instead of deriving phase boundaries directly from the PBC spectrum, we use the argument principle to reformulate the problem as counting the poles and zeros of a complex function defined from the PBC Hamiltonian, following the standard formulation of non-Hermitian winding numbers.~\cite{Okuma:2020} We then combine the Schur-Cohn method~\cite{Schur:1917,Cohn:1922,Henrici:1974,Rahman:2002,Stoica:1992} with the resultant method~\cite{Ahlfors:1979,Lang:2013,Cox:1997,Gelfand:1994} to obtain winding numbers and analytic phase boundaries.
 
We apply this approach to two representative 1D NHTSC lattices, with and without sublattices, which extend Refs.~\citenum{Chang:2026,Okuma:2019} to a larger parameter space. The resulting winding numbers and analytic phase boundaries are consistent with the number of MZMs obtained from numerically solved OBC spectra when the NHSE is suppressed by a weak onsite transverse magnetic field.
Beyond applying the Schur-Cohn method to the present analysis, we also discuss the role of the perturbation in  the symmetries and topology. In particular, we distinguish winding numbers associated with NHSE from those associated with MZMs and clarify their mutual relations before and after adding the perturbation. We further verify the suppression of the NHSE through bulk-state density profiles, finite-size comparisons between OBC and PBC spectra, and the average inverse participation ratio (IPR) of the (right) eigenstates under OBC.
To demonstrate the stability and identification of the MZMs, we analyze the finite-size scaling of the near-zero-mode energies and their IPRs, and examine their robustness against onsite potential disorder. In particular, we perform disorder averaging of the zero-mode splittings, bulk gap, and zero-mode IPRs over multiple disorder realizations, and clarify that the disorder term preserves the relevant internal symmetries. We also show that phases with $|W|=3$ can be obtained by including third-nearest-neighbor hopping processes, with the corresponding OBC spectra, boundary-mode profiles, and finite-size scaling supporting the emergence of three pairs of MZMs. These results establish a general analytic framework for higher-winding 1D non-Hermitian topological superconductors.

The rest of the article is organized as follows. In Sec.~\ref{Sec:General}, we provide an overview of the point-gap winding number for 1D non-Hermitian systems and the Schur-Cohn method for evaluating the winding number, as well as deducing the analytical expression for phase boundaries. In Sec.~\ref{Sec:NHTSC}, we consider a 1D NHTSC model with sublattice,  and examine its symmetries, the winding number responsible for the NHSE and the one associated with the MZMs, topological phase diagrams, the corresponding spectral and eigenstate properties, and the robustness of MZMs against onsite disorder. We also verify the correspondence between the winding numbers computed with and without the perturbation.
We further
demonstrate an even higher-winding phase with $W=3$ by including the third-nearest-neighbor hopping terms. 
In Sec.~\ref{sec:NHTSC2}, we apply the Schur-Cohn method to another 1D NHTSC model, but without sublattice, to demonstrate the broad applicability of this method. 
Finally, we present the discussion and conclusions in Sec.~\ref{Sec:Discussion}.
We provide details about computing the winding number based on the Schur-Cohn method  in Appendix~\ref{Appendix:Schur-Cohn_detail} and details about analytically deriving the phase boundaries   in Appendix~\ref{Appendix:deducing_phase_boundaries}. 
Details about the analysis of the first NHTSC model are presented in Appendix~\ref{Appendix:Details_NHTSC}.
Additional numerical results are presented in Appendix~\ref{sec:additional_results} for completeness.

\section{1D non-Hermitian topological phases characterized by $\mathbb{Z}$ invariants } 
\label{Sec:General}

\subsection{Point gap winding number for general systems}
\label{Sec:Winding}

We consider general 1D non-Hermitian systems that host topologically nontrivial phases characterized by point-gap topology with $\mathbb{Z}$ invariant(s). Under the PBC, one can construct a Bloch Hamiltonian, $ \mathbb{H} (k)$, with the crystal momentum $k$. Assuming that the  1D PBC Hamiltonian is irreducible with a point gap, we can define the winding number, 
\begin{align}
\mathbb{W} (E_{\rm ref})
=
\int^{2\pi/a_0}_0
\frac{dk}{2\pi i}\,
\frac{d}{dk}
\ln \big\{ \det[ \mathbb{H} (k) - E_{\rm ref} \mathbb{I} ]\big\},
\label{Eq:Winding_general_definition}
\end{align} 
with the reference energy $  E_{\rm ref}$, the identity matrix $\mathbb{I}$, and the lattice constant $a_0$. If $ \mathbb{H} (k)$ is reducible, it can be block-diagonalized into irreducible blocks, and a similar winding number can be defined for each block. 
Geometrically, Eq.~\eqref{Eq:Winding_general_definition} counts how many times $\det[\mathbb{H}(k)-E_{\rm ref} \mathbb{I}]$ winds around the origin as $k$ spans the Brillouin zone, $k \in (-\pi/a_0,\pi/a_0]$. It also identifies possible PBC point-gap closings: $\mathbb{W}$ can change only when the determinant vanishes for some $k$, namely when the PBC spectrum touches the reference energy $E_{\rm ref}$. Thus, $\mathbb{W}(E_{\rm ref})$ is an integer topological invariant protected by the point gap at $E_{\rm ref}$. For $E_{\rm ref}=0$, this corresponds to a point gap at zero energy.

However, a direct evaluation of Eq.~\eqref{Eq:Winding_general_definition} can become impractical for systems with large internal degrees of freedom or more elaborate lattice structures, such as longer-range hoppings. We therefore adopt the Schur-Cohn method, which provides a systematic way to determine the relevant root distribution and hence compute the winding number for general PBC Hamiltonians.

\subsection{Evaluation of the winding number}
\label{Sec:Winding_Evaluation}

We now introduce our approach to evaluate the winding number through the argument principle and Schur-Cohn method. 
Away from the phase transition, the determinant in Eq.~\eqref{Eq:Winding_general_definition} is nonzero for all real $k$, so that the winding number is well defined.
 Introducing the complex variable $z=e^{ik a_0}$,  the Brillouin zone is mapped to the unit circle \(|z|=1\) in the complex plane of $z$ 
and the determinant can be expressed as a polynomial  of \(z\),
\begin{equation}
\left. \det [ \mathbb{H} (k) - E_{\rm ref} \mathbb{I}] \right|_{e^{ik a_0} \to z }
=
\sum_{\ell=-d}^{d} C_{\ell} z^{\ell},
\label{Eq:Laurent_polynomial_det_zero}
\end{equation}
where the coefficients $C_{\ell}$ are complex functions of the Hamiltonian parameters. 
The integer \(d\) denotes the order of the pole of the above polynomial at the origin. 
By multiplying with $z^d$, this pole can be removed, and one obtains an ordinary complex polynomial, 
\begin{equation}
P_{2d} (z) 
\equiv \sum_{\ell=-d}^{d} C_{\ell} z^{d + \ell}.
\label{Eq:ordinary_polynomial_from_Laurent}
\end{equation}

Since the term with \(z^{-d}\) in Eq.~\eqref{Eq:Laurent_polynomial_det_zero} contributes a pole of order \(d\), applying the argument principle to Eq.~\eqref{Eq:Winding_general_definition} gives 
\begin{equation}
\mathbb{W}
=
\mathbb{N}[ P_{2d} (z) ]-d,
\label{Eq:winding_argument_principle_zero}
\end{equation}
where \( \mathbb{N}[P_{2d} (z)]\) denotes the number of zeros of \(P_{2d} (z)\) inside the unit circle, \(|z|<1\).  
Therefore, the computation of the winding number is reduced to the problem of counting the number of zeros of a complex polynomial inside the unit circle, a procedure also demonstrated in, e.g., Refs.~\citenum{Okuma:2020,Okuma:2023}.

Naively, one may attempt to evaluate Eq.~\eqref{Eq:winding_argument_principle_zero}
by solving the polynomial equation $P_{2d}(z)=0$. However, for high-degree polynomials with $2d>4$, no general closed-form solution for the roots exists. 
Thus, solving the polynomial equation would generally require numerical root finding, which does not provide an analytic alternative to the direct numerical evaluation of the winding integral in Eq.~\eqref{Eq:Winding_general_definition}. 

To avoid this difficulty, we adopt the Schur-Cohn method from the literature on polynomial theory and control systems,~\cite{Schur:1917,Cohn:1922,Henrici:1974,Rahman:2002,Stoica:1992} which determines the number $ \mathbb{N}$ directly from the coefficients of a complex polynomial, without explicitly solving for its roots.
In practice, our method involves a recursive procedure,   
with the starting point given by Eq.~\eqref{Eq:ordinary_polynomial_from_Laurent}. 
In the first step,
we compare the magnitudes of the leading-term coefficient \(C_{d}\) and the constant coefficient \(C_{-d}\). 
If $|C_{d}|>|C_{-d}|$, we construct the reduced polynomial,
\begin{eqnarray}
    P^{\rm red}_{2d-1} (z)
&\equiv&
\frac{C_{d}^*\,P_{2d}(z)-C_{-d}\,\overline{P}_{2d}(z)}{z} \nonumber\\
&=& 
\sum_{\ell=-d}^{d-1}
\left(
C_{d}^* C_{\ell+1}
-
C_{-d}\,C_{-\ell-1}^*
\right) z^{ d + \ell } ,
\end{eqnarray} 
where we define the reciprocal polynomial of \(P_{2d}(z)\),
\begin{equation}
\overline{P}_{2d}(z)= z^{2d} [ P_{2d} (1/z^*) ]^*. 
\end{equation}
If, instead, we have $|C_{d}| < |C_{-d}|$, we use the reciprocal relation, 
\begin{equation}
\mathbb{N}[P_{2d} (z) ]=2d- \mathbb{N}[\overline{P}_{2d} (z) ],
\end{equation}
and continue the same recursive procedure for \(\overline{P}_{2d} (z)\) to construct the corresponding reduced polynomial. 

At each step, this operation lowers the polynomial degree while keeping track of how the zero count of the original polynomial is related to that of the reduced polynomial.
As detailed in Appendix~\ref{Appendix:Schur-Cohn_detail}, by repeating this procedure until the polynomial is reduced to a nonzero constant, one obtains the final value of $\mathbb{N}$ and hence the winding number through Eq.~\eqref{Eq:winding_argument_principle_zero}.
 This coefficient-based implementation avoids explicit root finding and therefore enables efficient computation of the winding number over broad parameter space.

The above procedure assumes that the leading and constant coefficients have unequal magnitudes at each step, and the standard Schur-Cohn procedure is applicable.~\cite{Schur:1917,Cohn:1922,Henrici:1974,Rahman:2002,Stoica:1992} When this condition is not satisfied, the recursion does not directly determine the winding number. Such points correspond to candidate phase boundaries, where zeros of the relevant polynomial may cross the unit circle and the winding number can change. 
As a result, the phase boundaries can also be identified.
In the following section, we discuss this case in more detail.

\subsection{Analytical expression for  phase boundaries }
\label{Sec:Deducing_phase_boundaries}

In addition to computing the winding number itself, the polynomial derived above can also be used to deduce the analytical expression for  phase boundaries in terms of the system parameters. Since the winding number is determined by the number of zeros of $P_{2d}(z)$ inside the unit circle, a zero crossing the unit circle changes $\mathbb{N}[P_{2d}(z)]$ and hence changes the winding number. Such a change occurs when $P_{2d}(z)$ has a zero on the unit circle, $z_0=e^{ik_0 a_0}$. Equivalently, the PBC spectrum touches the reference energy, 
$ \det[ \mathbb{H} (k_0) - E_{\rm ref} \mathbb{I} ]=0$, 
 signaling the closing of the point gap at $E_{\rm ref}$. For $E_{\rm ref}=0$, this reduces to a zero-energy PBC gap closing. Therefore, the phase boundary is determined by
\begin{equation}
P_{2d}(z_0)=0,\qquad |z_0|=1 .
\label{Eq:phase_boundary_general}
\end{equation}

Since $z_0=1/z_0^{*}$ on the unit circle, a zero of $P_{2d}(z)$ on the unit circle is also a zero of its reciprocal polynomial $\overline{P}_{2d}(z)$. Thus, a PBC gap closing implies that $P_{2d}(z)$ and $\overline{P}_{2d}(z)$ have a common zero. Without explicitly solving for $k_0$, this condition can be imposed through the resultant
\begin{equation} 
    \mathrm{Res}_z\!\left[P_{2d}(z),\overline{P}_{2d}(z)\right],
    \label{Eq:phase_boundary_half-analytic}
\end{equation}
which allows us to obtain analytical expressions for candidate phase boundaries by setting the resultant to zero. The condition may contain extra solutions with $|z_0|\neq 1$, for instance from reciprocal root pairs of $P_{2d}(z)$ not on the unit circle.
The genuine conditions of the phase boundaries can then be pinned down by comparing with the winding-number calculation described in Sec.~\ref{Sec:Winding_Evaluation} and retaining only those parameter values across which the winding number changes. 
A fully analytical way to obtain the genuine phase boundaries, without combining with numerical calculations, is possible by recasting the
condition in Eq.~\eqref{Eq:phase_boundary_general} into a real-polynomial problem, as detailed
in Appendix~\ref{Appendix:deducing_phase_boundaries}.

Before proceeding, we remark that the zeros of the polynomial in Eq.~\eqref{Eq:ordinary_polynomial_from_Laurent} are written in terms of the same complex variable $z=e^{ik a_0}$ that also appears in non-Bloch band theory. In the latter context, one extends $z$ away from the unit circle and interprets it as a non-Bloch momentum variable under OBC.~\cite{Yao:2018,Yokomizo:2019,Joshi:2026} 
In the present work, however, these zeros are not introduced as non-Bloch momenta or as a direct diagnostic of OBC boundary modes. Their primary role is instead to provide an analytic criterion for locating the PBC phase boundaries, where the relevant condition is that a zero crosses the unit circle $|z|=1$.
This PBC criterion should therefore be distinguished from the non-Bloch bulk-boundary correspondence, which relates bulk topology to OBC boundary spectra through a generalized non-Bloch description in non-Hermitian systems.~\cite{Yao:2018-2,Ghosh:2022}  
This distinction between the present polynomial method and the earlier non-Bloch analysis is also relevant to the NHSE.
In non-Bloch band theory this behavior is captured by replacing the unit circle of the ordinary Brillouin zone with a generalized Brillouin zone.~\cite{Yao:2018,Yokomizo:2019} In contrast, the  analysis  here employs the polynomial zeros only to determine whether they cross the unit circle $|z|=1$, which gives the PBC gap-closing condition.

Having presented the general guideline for evaluating the winding number characterizing 1D non-Hermitian topological phases with $\mathbb{Z}$ invariants, below we provide concrete examples by examining two specific models.

\begin{figure}[h]
    \centering
    \includegraphics[width=0.48\textwidth]{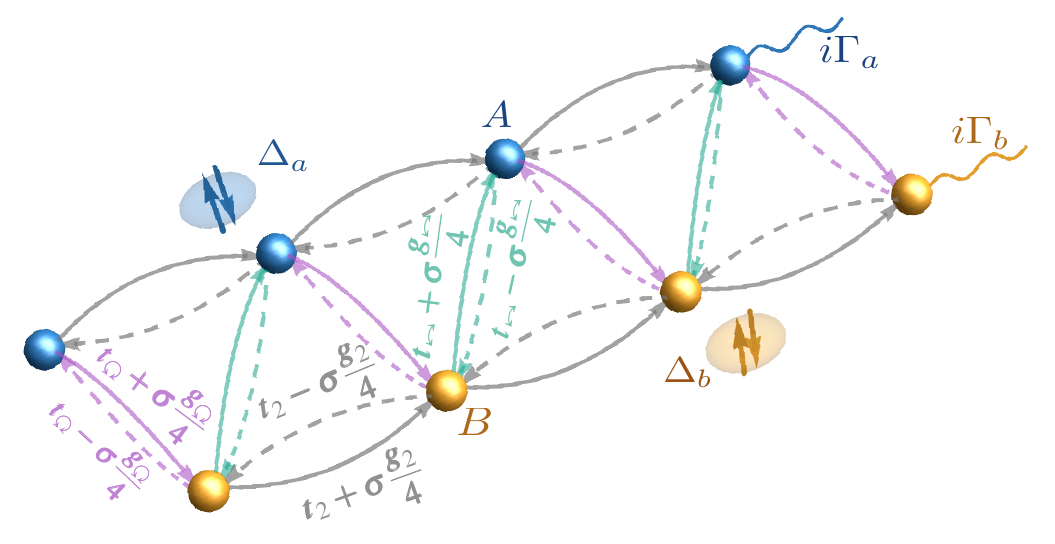} 
    \caption{
    Schematic illustration of the 1D lattice model described by Eq.~\eqref{Eq:H_NHTSC}, consisting of two sublattices labeled by $A$ and $B$. Arrows indicate different hopping processes, distinguished by colors and by solid or dashed lines. The onsite pairings  $\Delta_{a,b}$ and  dissipation $i\Gamma_{a,b}$ are indicated by ellipses and wavy lines, respectively.
    For clarity, the lattice is displayed in a zigzag configuration.  
     }
    \label{fig:schematic}
\end{figure}

\section{1D NHTSC model with sublattice}
\label{Sec:NHTSC}

\subsection{Hamiltonian and symmetry properties}
\label{sec:H_and_symm}

We  start with a 1D NHTSC model with sublattice, including nonreciprocal hopping, dissipation, and pairing terms,  generalized from Ref.~\onlinecite{Chang:2026}. As schematically depicted in Fig.~\ref{fig:schematic}, the real-space Hamiltonian is given by
\begin{widetext}
\begin{align}
H_{\mathrm{NHTSC}}
=&\sum_{j,\sigma}\Big[
-\left(t_{\inter}+\sigma\tfrac{g_{\inter}}{4}\right)\,a_{j+1,\sigma}^{\dagger}b_{j,\sigma}
-\left(t_{\inter}-\sigma\tfrac{g_{\inter}}{4}\right)\,b_{j,\sigma}^{\dagger}a_{j+1,\sigma}  
-\left(t_{\intra}-\sigma\tfrac{g_{\intra}}{4}\right)\,a_{j,\sigma}^{\dagger}b_{j,\sigma}
-\left(t_{\intra}+\sigma\tfrac{g_{\intra}}{4}\right)\,b_{j,\sigma}^{\dagger}a_{j,\sigma}  \nn
&\quad
-(t_{2}-\sigma\tfrac{g_{2}}{4})\,a_{j,\sigma}^{\dagger}a_{j+1,\sigma}
-(t_{2}+\sigma\tfrac{g_{2}}{4})\,a_{j+1,\sigma}^{\dagger}a_{j,\sigma} 
-(t_{2}-\sigma\tfrac{g_{2}}{4})\,b_{j,\sigma}^{\dagger}b_{j+1,\sigma}
-(t_{2}+\sigma\tfrac{g_{2}}{4})\,b_{j+1,\sigma}^{\dagger}b_{j,\sigma}
\Big] \nn
&-\frac{i}{2}\sum_{j,\sigma}\left(
\Gamma_a\,a_{j,\sigma}^{\dagger}a_{j,\sigma}
+\Gamma_b\,b_{j,\sigma}^{\dagger}b_{j,\sigma}
\right) +\sum_j\left(
\Delta_a\,a_{j,\uparrow}^{\dagger}a_{j,\downarrow}^{\dagger}
+\Delta_b\,b_{j,\uparrow}^{\dagger}b_{j,\downarrow}^{\dagger}
+\mathrm{H.c.}
\right),
\label{Eq:H_NHTSC}
\end{align}
\end{widetext}
where $a_{j,\sigma}$ ($b_{j,\sigma}$) annihilates a spin-1/2 fermion with spin $\sigma \in \{ \uparrow, \downarrow\}$ on sublattice $A$ ($B$) in a unit cell labeled by $j \in [1,N/2]$, where $N$ is the total number of lattice sites and is taken to be even throughout this work. 
The parameters $t_{\intra}$ and $g_{\intra}$ denote 
symmetric and antisymmetric components of the intra-cell nearest-neighbor hopping amplitudes, 
respectively. The parameters $t_{\inter}$ and $g_{\inter}$ describe their  inter-cell counterparts. The terms $t_2$ and $g_2$ represent the symmetric and antisymmetric components of the second-nearest-neighbor hopping within the same sublattice. 
In this model, non-Hermiticity is introduced through the sublattice-dependent onsite dissipation rates $\Gamma_{a,b}$ as well as the nonreciprocal hopping processes related to $ g_{\intra} $, $ g_{\inter} $,  and $ g_{2}$. 
The superconductivity is incorporated via on-site spin-singlet pairing with sublattice-dependent amplitudes $\Delta_{a,b}$.

Under the PBC, the Bogoliubov-de Gennes (BdG) form of the Hamiltonian is given by   
\[ 
{H}_{\mathrm{NHTSC}} = \frac{1}{2}  \sum_{k} \Psi_k^\dagger H^{\mathrm{pbc}}_{\mathrm{NHTSC}}(k)\, \Psi_k
\]
in the Nambu basis, $\Psi_k^\dagger =
\bigl(
a_{k,\uparrow}^\dagger,\,
a_{k,\downarrow}^\dagger,\,
b_{k,\uparrow}^\dagger,\,
b_{k,\downarrow}^\dagger,\,
a_{-k,\uparrow},\,
a_{-k,\downarrow},\,
b_{-k,\uparrow},\,
b_{-k,\downarrow}
\bigr)$,  with fermion fields, $a_{k,\sigma}$ and $b_{k,\sigma}$, in the momentum space and the Hamiltonian, 
\begin{widetext}
    \begin{align}
        H_{\rm NHTSC}^{\rm pbc}(k)
&= t_{\inter}\sin(ka_0)\,\eta^{z}\tau^{y}-\big[t_{\inter}\cos(ka_0)+t_{\intra}\big]\eta^{z}\tau^{x} -i\Big[\tfrac{g_{\inter}}{4}\cos(ka_0)-\tfrac{g_{\intra}}{4}\Big]\tau^{y}\sigma^{z}
-i\tfrac{g_{\inter}}{4}\sin(ka_0)\tau^{x}\sigma^{z}   \nn
&\quad  -2t_2\cos(ka_0)\,\eta^{z}+i\tfrac{g_2}{2}\sin(ka_0)\sigma^{z} -\frac{i}{2}\Gamma_{+}\eta^{z}-\frac{i}{2}\Gamma_{-}\eta^{z}\tau^{z} -\Delta_{+}\eta^{y}\sigma^{y}
-\Delta_{-}\eta^{y}\tau^{z}\sigma^{y}.
\label{Eq:H_pbc}
    \end{align}
\end{widetext}
In the above, $\eta^\mu$, $\tau^\mu$, and $\sigma^\mu$ are Pauli matrices acting on particle--hole, sublattice, and spin degrees of freedom, respectively, and we define $\Gamma_{\pm}=(\Gamma_a\pm\Gamma_b)/2$ and $\Delta_{\pm}=(\Delta_a\pm\Delta_b)/2$. In addition, for numerical convenience, we define the symmetric and antisymmetric combinations of the reciprocal nearest-neighbor hoppings as $t_{\pm} = (t_{\inter} \pm t_{\intra})/2$ and the nonreciprocal ones as $g_{\pm} = (g_{\inter} \pm g_{\intra})/2$.
The analysis in Ref.~\onlinecite{Chang:2026} corresponds to the regime of $t_2=g_2=0$ and mainly focuses on the limit of $\Gamma_{-}= \Delta_{-}=0$. 
Here we explore a much larger parameter space with nonzero $t_2$, $g_2$, $\Gamma_{-}$, and $\Delta_{-}$.
The full analytical expression for the PBC spectrum, given in Eq.~\eqref{Eq:PBC_spec_general_case} in Appendix~\ref{Appendix:lengthy}, is rather complex, making a direct derivation of the gap-closing conditions and phase boundaries difficult.
 
Following Refs.~\onlinecite{Okuma:2019,Chang:2026}, we introduce a perturbation term in the form of a weak onsite transverse magnetic field,
 \begin{align}
    H_{\rm pt}  = \delta h_x \eta^z  \sigma^x,
    \label{eq:pert_field}
\end{align}
which  suppresses the NHSE while preserving the MZMs, if present. It is therefore meaningful to examine the OBC spectra of $H_{\mathrm{NHTSC}}+H_{\rm pt}$ and compare them with the winding number computed from Eq.~\eqref{Eq:H_pbc} under the PBC. 
Before computing the winding number, we discuss the symmetry properties of the model, which depend on whether the perturbation in Eq.~\eqref{eq:pert_field} is included. Notably, the second-nearest-neighbor hopping terms break some of the internal symmetries present in the model of Ref.~\citenum{Chang:2026}, while preserving the important symmetry responsible for the presence of MZMs, as discussed in Sec.~\ref{sec:topo-inv}.

We first consider the regime without the perturbation. In this case, $H^{\rm pbc}_{\rm NHTSC}$ possesses a nontrivial unitary symmetry, represented by $U = \eta^z\tau^0\sigma^z$, and is therefore block diagonalizable. As summarized in Table~\ref{Table:Symmetries_without_perturbation}, it also preserves particle-hole symmetry (PHS), time-reversal-dagger symmetry (TRS$^\dagger$), and sublattice symmetry (SLS), represented by $U_{C_-}$, $U_{C_+}$, and $U_S$, respectively.
The full symmetry group is $\mathbb{Z}_2^{\times 3}$, which can be generated by $U$ together with any two of $U_{C_-}$, $U_{C_+}$, and $U_S$.

To proceed, we block diagonalize $H^{\rm pbc}_{\rm NHTSC}$ and analyze the symmetry classification of its irreducible blocks. Specifically, we introduce a unitary matrix $V$ satisfying
$V^\dagger U V={\rm diag}(\mathbb{I}_{4\times4},-\mathbb{I}_{4\times4})$, under which the Hamiltonian becomes
\begin{align}
    V^\dagger H^{\rm pbc}_{\rm NHTSC}(k) V
    =
    \begin{pmatrix}
        H_+(k) & 0 \\
        0 & H_-(k)
    \end{pmatrix} ,
    \label{eq:V_H_V}
\end{align} 
with  the corresponding  blocks, $H_\pm$, in the $\pm1$ eigenspaces of $U$.
We find that both reduced blocks, $ H_{\pm} (k)$, possess only SLS; see Appendix~\ref{sec:App_symm}. Accordingly, they belong to class A with SLS in the complex AZ classification, which supports a $\mathbb{Z}\oplus\mathbb{Z}$ topological invariant in the point-gap case and a $\mathbb{Z}$ topological invariant in the line-gap case.
 
When Eq.~\eqref{eq:pert_field} is introduced into $H^{\rm pbc}_{\rm NHTSC}$, the up- and down-spin sectors become coupled, and the unitary symmetry is broken. 
As a result, the remaining symmetry group is reduced to $\mathbb{Z}_2^{\times 2}$, generated by any two of PHS, TRS$^\dagger$, and SLS, as shown in Eq.~\eqref{eq:Symm_pert_H_dhx} of Appendix~\ref{sec:App_symm}. 
The resulting symmetry structure admits two equivalent descriptions: class D with an additional $S_+$-type SLS in the real AZ classification, and class AI$^\dagger$ with the same additional SLS in the real AZ$^\dagger$ classification. 
Here, $S_+$ denotes that the SLS commutes with PHS. 
This symmetry class admits a $\mathbb{Z}$ topological invariant in both the point-gap and line-gap classifications.

Having discussed how the perturbation changes the topological class, we now show how the irreducible blocks built from $H^{\rm pbc}_{\rm NHTSC}$ and $H^{\rm pbc}_{\rm NHTSC}+H_{\rm pt}$ are connected, which is important for constructing the winding numbers in the two cases.

\subsection{Symmetry-constrained forms for the irreducible blocks
}
\label{sec:H_reduction}

In this section, we discuss how the perturbation term affects the symmetries of the model and, consequently, the structure of the irreducible blocks. The discussion applies to general system parameters. 
Since the SLS operator $U_S$ is proportional to the product of $U_{C_+}U_{C_-}$ up to a phase factor, only two of these operators are independent. In the following discussion, we therefore focus on $U_{C_+}$ and $U_S$, together with the operator $U$ when the unitary symmetry is preserved.

We first discuss the system in the absence of the perturbation.  
Since the nontrivial unitary operator $U$ commutes with both $H^{\rm pbc}_{\rm NHTSC}$ and $U_S$, in the basis that diagonalizes $U$, both $H^{\rm pbc}_{\rm NHTSC}$ and $U_S$ take block-diagonal forms. Specifically, with the same $V$ that leads to   Eq.~\eqref{eq:V_H_V}, we have  
     \begin{eqnarray}
     V^\dagger U_S  V
     & =
    \begin{pmatrix}
        u_{S,+} & 0 \\
        0 & u_{S,-}
    \end{pmatrix},
    \label{eq:H_V_Bdiag}
    \end{eqnarray} 
where  $u_{S,\pm}$ are the corresponding $4\times4$ blocks in the $\pm1$ eigenspaces of $U$. 

The SLS   then implies
\begin{align}
    u_{S,\pm} H_\pm(k) u_{S,\pm}^\dagger
    =
    -H_\pm(k),
    \label{eq:relation_H_uS}
\end{align}
and that  each of the blocks, $H_\pm $, also possesses SLS.  The anticommutation relation in Eq.~\eqref{eq:relation_H_uS} further forces the diagonal blocks of $H_\pm(k)$ to vanish, yielding the chiral form in the eigenbasis of $u_{S,\pm}$: 
\begin{align}
\label{eq:Z_H_Z}
   & \begin{pmatrix}
        Z_+^\dagger & 0 \\
        0 & Z_-^\dagger
    \end{pmatrix}   \begin{pmatrix}
        H_+(k) & 0 \\
        0 & H_-(k)
    \end{pmatrix}  \begin{pmatrix}
        Z_+ & 0 \\
        0 & Z_-
    \end{pmatrix}  \nn
= &
\begin{pmatrix}
0 & h_+(k) & 0 & 0\\
h'_+(k) & 0 & 0 & 0\\
0 & 0 & 0 & h_-(k)\\
0 & 0 & h'_-(k) & 0
\end{pmatrix},
\end{align}
with  $Z_\pm$ the unitary matrices that diagonalize $u_{S,\pm}$ such that $Z_\pm^\dagger u_{S,\pm} Z_\pm = {\rm diag}(\mathbb{I}_{2\times2},-\mathbb{I}_{2\times2})$, and $h_\pm(k)$ and $h_\pm^\prime(k)$ are the $2\times2$ blocks arising from the transformation of $H_{\pm}(k)$.
The TRS$^\dagger$, together with the relations $U_{C_+} U^* = -U U_{C_+}$ and $U_{C_+} U_S^* = U_S U_{C_+}$, further constrains $h_\pm(k)$ and $h_\pm^\prime(k)$, yielding
\begin{subequations}
\label{eq:hpm_hpmPrime}
    \begin{eqnarray}
         h_+^\prime(k) &= h_-^T(-k), \\
    h_-^\prime(k) &= h_+^T(-k), 
    \end{eqnarray}
\end{subequations}
with the transpose operator $T$. 
Therefore, only two of the four blocks  are independent, from which we construct the subsystem winding numbers, as shown in Sec.~\ref{sec:topo-inv}.
One can further reorder the basis according to $\mathcal P (\varphi_{+,+},\, \varphi_{+,-},\, \varphi_{-,+},\,\varphi_{-,-}) = (\varphi_{+,+},\, \varphi_{-,+},\, \varphi_{+,-},\,\varphi_{-,-}) 
$, and get 
\begin{align}
    & \mathcal{P}\begin{pmatrix}
0 & h_+(k) & 0 & 0\\
h'_+(k) & 0 & 0 & 0\\
0 & 0 & 0 & h_-(k)\\
0 & 0 & h'_-(k) & 0
\end{pmatrix}
\mathcal{P}^\dagger \nn
=& \begin{pmatrix}
0 & 0 & h_+(k) & 0\\
0 & 0 & 0 & h_-(k)\\
h'_+(k) & 0 & 0 & 0\\
0 & h'_-(k) & 0 & 0
\end{pmatrix}.
\label{Eq:HhatW_SLS_grouped_order}
\end{align}
Here, $\varphi_{\pm,\pm}$ denotes the basis states after the transformation $V{\rm diag}(Z_+,Z_-)$, with the first index labeling the corresponding sector to the eigenvalue $(=\pm 1)$ of $V$ and the second index labeling that of  ${\rm diag}(Z_+,Z_-)$. 
It will be shown below that the form of Eq.~\eqref{Eq:HhatW_SLS_grouped_order} is useful when comparing the irreducible blocks before and after introducing the perturbation. 

Next, we consider the effect of the perturbation $H_{\rm pt}$ given in Eq.~\eqref{eq:pert_field}. This perturbation breaks the nontrivial unitary symmetry $U$, while preserving SLS and TRS$^\dagger$. Thus, at the level of internal symmetries, we have 
\begin{align}
   {\rm Unitary} +{\rm SLS}+ {\rm TRS}^\dagger
   \xrightarrow{H_{\rm pt}}
   {\rm SLS} + {\rm TRS}^\dagger .
\end{align}
In a basis where the SLS operator $U_S$ is diagonal, we have 
\begin{subequations}
\label{Eq:HhatW_V_zero_perturbation} 
    \begin{eqnarray}
       V_{\rm pt}^\dagger \Big( H^{\rm pbc}_{\rm NHTSC} (k) + H_{\rm pt} \Big) V_{\rm pt}
& =&
\begin{pmatrix}
0 & M(k)\\
M^{\prime}(k) & 0
\end{pmatrix}, 
\label{eq:R_Hpt_R} 
\\
V_{\rm pt}^\dagger  U_{C_+} V^*_{\rm pt}
&  = &
\begin{pmatrix}
u_C & 0 \\
0 & u_C^\prime
\end{pmatrix},
    \end{eqnarray}
\end{subequations}
where $u_C$ and $u_C^\prime$ are symmetric and unitary, and  $V_{\rm pt}$ is the unitary matrix that diagonalizes  $U_S$. The TRS$^\dagger$ further relates the blocks, $M(k)$ and $M^\prime (k)$, through the relation
\begin{align}
    M^\prime (-k) = u_{C}^\prime M^T (k) u_{C}^\dagger.
    \label{eq:V_prime_V}
\end{align}
As summarized in Eqs.~\eqref{Eq:HhatW_V_zero_perturbation}--\eqref{eq:V_prime_V}, the SLS  allows one to transform the full Hamiltonian into an off-diagonal form, while the TRS$^\dagger$  relates these off-diagonal components.

As shown above, the basis transformations through $V$ and $V_{\rm pt}$ place the unperturbed and perturbed Hamiltonians in qualitatively different forms. The unperturbed Hamiltonian becomes block diagonal in the transformed basis [Eq.~\eqref{eq:Z_H_Z}], whereas the perturbed one takes an off-block-diagonal form [Eq.~\eqref{eq:R_Hpt_R}] due to the absence of the nontrivial unitary symmetry.
Comparing Eq.~\eqref{eq:R_Hpt_R} to Eq.~\eqref{Eq:HhatW_SLS_grouped_order}, one finds that $M(k)$ and $M^\prime(k)$ reduce to the two off-diagonal blocks appearing in the right-hand side of Eq.~\eqref{Eq:HhatW_SLS_grouped_order} in the unperturbed limit.
The additional transformation in Eq.~\eqref{Eq:HhatW_SLS_grouped_order} therefore brings the unperturbed Hamiltonian into a form better suited for comparison with the perturbed one and for the subsequent analysis of the topological invariants.

The relations between the irreducible blocks before and after adding the perturbation indicate a close connection between the topological invariants of $H^{\rm pbc}_{\rm NHTSC}$ and that of $H^{\rm pbc}_{\rm NHTSC}+H_{\rm pt}$, as we discuss next.

\subsection{Subsystem and composite winding numbers   }
\label{sec:topo-inv}

In this section, we discuss the winding numbers in both the absence and presence of the perturbation, focusing on their roles as topological invariants characterizing the NHSE and topological zero-energy modes.

We begin with the unperturbed case. 
Motivated by the off-diagonal form in Eq.~\eqref{eq:Z_H_Z}, we define the winding numbers around the origin in the complex energy plane associated with the four $2\times 2$ blocks as
\begin{subequations}
\label{eq:w_pm_prime}
\begin{eqnarray}
w_\pm
    &=
    \int_0^{2\pi/a_0}
    \frac{dk}{2\pi i}\,
    \frac{d}{dk}
    \ln
    \left\{
        \det\bigl[h_\pm(k)\bigr]
    \right\}, \\
w_\pm^\prime
    &=
    \int_0^{2\pi/a_0}
    \frac{dk}{2\pi i}\,
    \frac{d}{dk}
    \ln
    \left\{
        \det\bigl[h^\prime_\pm(k)\bigr]
    \right\}.
\end{eqnarray}
\end{subequations} 
As discussed in Sec.~\ref{Sec:General}, the determinants above can be expressed in terms of complex polynomials of $z = e^{i k a_0}$ as in Eq.~\eqref{Eq:Laurent_polynomial_det_zero}.
Then, by introducing the corresponding ordinary polynomials as in Eq.~\eqref{Eq:p(z)_app}, we follow the procedure in Appendix~\ref{Appendix:Schur-Cohn_detail} and Appendix~\ref{Appendix:deducing_phase_boundaries} to compute the corresponding winding numbers.

From Eq.~\eqref{eq:hpm_hpmPrime}, only two of these winding numbers are independent, satisfying
\begin{equation}
    w_\pm^\prime = -\,w_\mp.
    \label{eq:dependence_w}
\end{equation}
As shown below, suitable combinations of these winding numbers characterize the NHSE and the emergence of MZMs.
Specifically, owing to the block-diagonal structure in Eq.~\eqref{eq:V_H_V}, one can define the winding numbers separately for the  sectors $H_\pm$  as 
\begin{align}
    \Wnhse_\pm (E_{\rm ref})
    &=
    \int_0^{2\pi/a_0}
    \frac{dk}{2\pi i}\,
    \frac{d}{dk}
    \ln
    \left\{
        \det\bigl[H_\pm(k) - E_{\rm ref} \mathbb{I}_{4 \times 4} \bigr]
    \right\}, 
\end{align}
with the reference energy $E_{\rm ref}$. 
These winding numbers are related to the emergence of the NHSE under the OBC. 
As discussed above, since the TRS$^\dagger$ connects the two sectors labeled by the $\pm$ signs through momentum inversion, $k\to -k$, we find 
\begin{align}
    \Wnhse_+  (E_{\rm ref}) = - \Wnhse_-  (E_{\rm ref}),
    \label{eq:W_nhse_pm}
\end{align}
indicating that only one of $\Wnhse_\pm$ remains independent.
The NHSE thus appears when $\Wnhse_\pm$ is nonzero for a reference energy inside a point gap. In terms of $w_\pm$ and $w_\pm^\prime$, which are defined for $E_{\rm ref}=0$, we find
\begin{align}
      \Wnhse_+ (0) = w_+ + w_+^\prime = w_+ - w_- \equiv \Delta W,
    \label{Eq:W_nhse}
\end{align}
where we have used the relation in Eq.~\eqref{eq:dependence_w}. 

On the other hand, motivated by the chiral form of Eq.~\eqref{Eq:HhatW_SLS_grouped_order}, we define a winding number associated with its upper-right block, diag ($h_+,h_-$), which gives
\begin{align}
    W =  w_+ + w_- .
   \label{eq:W_nonHermi} 
\end{align}
 Consequently, the unperturbed system is characterized by a $\mathbb{Z}\oplus\mathbb{Z}$ topological invariant, represented by $W$ and $\Wnhse_+$, consistent with the discussion in Sec.~\ref{sec:H_and_symm}.
 The topological invariant $W$ suggests the emergence of boundary zero modes. However, owing to the NHSE, the OBC and PBC spectra differ significantly, making the boundary zero modes difficult to identify directly from the OBC spectra.

This motivates the introduction of the perturbation
in Eq.~\eqref{eq:pert_field}. In this case, the full Hamiltonian $H^{\rm pbc}_{\rm NHTSC}+H_{\rm pt}$ can no longer be decomposed into two independent sectors by the unitary symmetry. The invariant associated with the NHSE must therefore be defined from the total winding number of the full Hamiltonian. The perturbation hybridizes the $\pm$ sectors in Eq.~\eqref{eq:V_H_V}, while the remaining TRS$^\dagger$ enforces the cancellation of their total winding over the entire complex-energy plane. As a result, the total winding number vanishes for any reference energy, leading to the suppression of the NHSE. 

We are then in a position to construct the topological invariant characterizing the MZMs. To this end, we note that, since SLS is preserved, the full Hamiltonian can still be transformed into the chiral form in Eq.~\eqref{Eq:HhatW_V_zero_perturbation}. Its topology is therefore characterized by the winding number of the off-diagonal block, $M(k)$, in Eq.~\eqref{eq:R_Hpt_R}, 
\begin{align}
    W^{\rm mzm}
    &=
    \int_0^{2\pi/a_0}
    \frac{dk}{2\pi i}\,
    \frac{d}{dk}
    \ln
    \left\{
        \det\bigl[M(k)\bigr]
    \right\} ,
    \label{eq:W_by_V}
\end{align}
as a $\mathbb{Z}$ invariant. This winding number can be connected to the unperturbed limit, in which the chiral form in Eq.~\eqref{Eq:HhatW_V_zero_perturbation} reduces to Eq.~\eqref{Eq:HhatW_SLS_grouped_order}, and the winding number in Eq.~\eqref{eq:W_by_V} reduces back to 
 Eq.~\eqref{eq:W_nonHermi}. 
 Thus, Eq.~\eqref{eq:W_by_V} provides the natural extension of the invariant to the perturbed case, while reproducing the unperturbed result when the perturbation strength is sufficiently weak.
In this weak-perturbation regime, we thus have $W^{\rm mzm} \to W$, and the invariant in Eq.~\eqref{eq:W_nonHermi} effectively characterizes the topological phases, as we demonstrate in the following section.

Based on the above symmetry-based analysis of the systems, we conclude that the unitary symmetry and TRS$^\dagger$, which exchanges the irreducible sectors of $U = \pm 1$,  are relevant to the appearance or suppression of the NHSE. In contrast, the SLS is related to the emergence of zero-energy modes that appear when the NHSE is suppressed. 

In the following section, we present numerical results for the topological phase diagrams for Eq.~\eqref{Eq:H_pbc}, based on the winding number defined here, along with the PBC and OBC spectra.

\begin{figure}[t]
    \centering
    \includegraphics[width=0.47\textwidth]{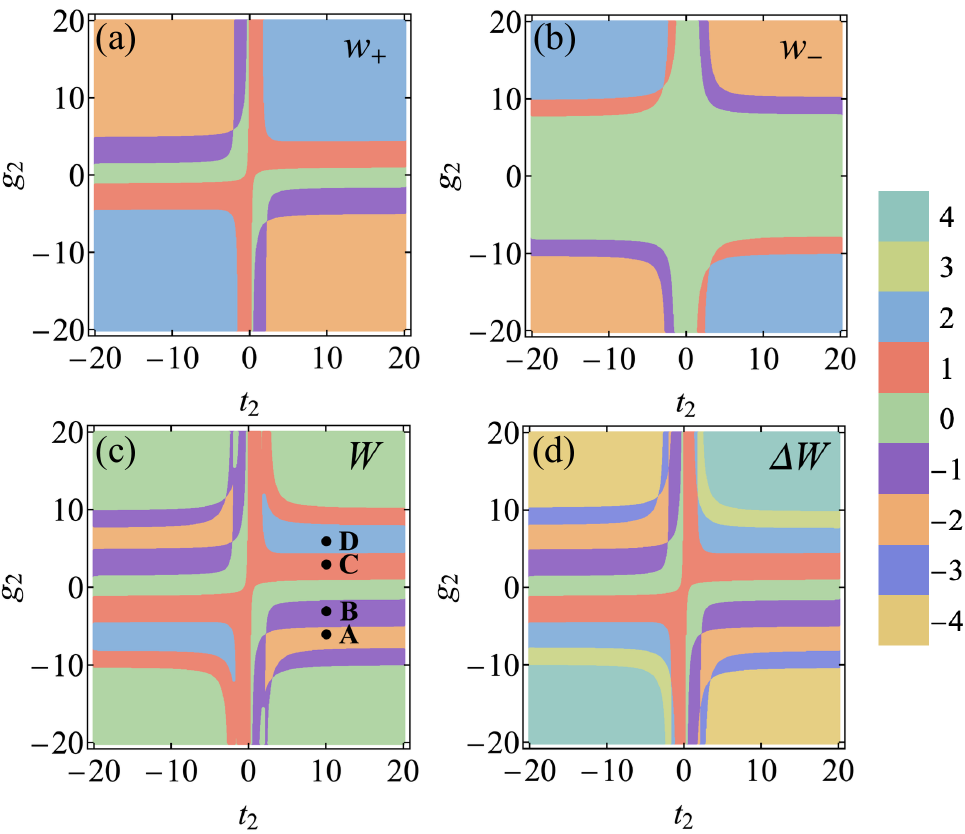}
    \caption{Phase diagrams based on the computed winding numbers in the $(t_2,g_2)$ plane. Panels (a) and (b) show $w_+$ and $w_-$, defined in Eq.~\eqref{eq:w_pm_prime}, corresponding to the blocks $h_+$ and $h_-$ given in Eq.~\eqref{eq:Z_H_Z}, respectively. Panels (c) and (d) show the composite winding numbers $W=w_+ + w_-$ and $ \Delta W = w_+ -w_-$, respectively. The adopted values of the remaining parameters are given by $4t_+=6$, $4t_-=10$, $g_+=3.0$, $g_-=3.5$, $\Gamma_+=3$, $\Gamma_-=3$, $\Delta_+=3$, and $\Delta_-=-1.5$.     
     }
    \label{fig:phase_diagram}
\end{figure}

\subsection{Topological phase diagrams and energy spectra}
\label{sec:PD_OBC}

In this section, we evaluate the winding numbers derived in the previous section and use them to construct the topological phase diagrams. As discussed above, it is sufficient to evaluate $w_\pm$ and $w_\pm^\prime$ defined in Eq.~\eqref{eq:w_pm_prime}, which are associated with the blocks $h_\pm(k)$ and $h_\pm^\prime(k)$ defined in Eq.~\eqref{eq:hpm_hpmPrime}. Using the Schur-Cohn method, we numerically compute $w_\pm$ and $w_\pm^\prime$, and then obtain the composite invariants $W$ and $\Delta W$ from the relations in Eqs.~\eqref{eq:W_nonHermi} and \eqref{Eq:W_nhse}, respectively.  

Representative results are shown in Fig.~\ref{fig:phase_diagram}. The winding numbers $w_\pm$ as a function of the second-nearest-neighbor hopping parameters are shown in Figs.~\ref{fig:phase_diagram}(a,b), along with the composite invariants $W$ and $\Delta W $   in Figs.~\ref{fig:phase_diagram}(c,d). 
We check that the phase boundaries in Fig.~\ref{fig:phase_diagram} are consistent with the analytical expression derived from the method introduced in Sec.~\ref{Sec:Deducing_phase_boundaries}. 
When longer-range hopping processes are included, topological phases with higher winding numbers, $|W|>1$, appear in certain parameter regimes, reflecting the underlying $\mathbb{Z}$ topology. 
To demonstrate the correspondence between the unperturbed winding number $W$ and the perturbed winding number $W^{\rm mzm}$, we construct the phase diagram in the presence of a weak onsite transverse perturbation by computing $W^{\rm mzm}$ in Eq.~\eqref{eq:W_by_V}. As shown in Fig.~\ref{fig:PD_dhx}(a) in Appendix~\ref{sec:additional_results}, the resulting phase diagram reproduces all phases of the unperturbed phase diagram in Fig.~\ref{fig:phase_diagram}, with only slight shifts of the phase boundaries, as expected for weak $\delta h_x$.

\begin{figure}[t]
    \centering
    \includegraphics[width=0.47\textwidth]{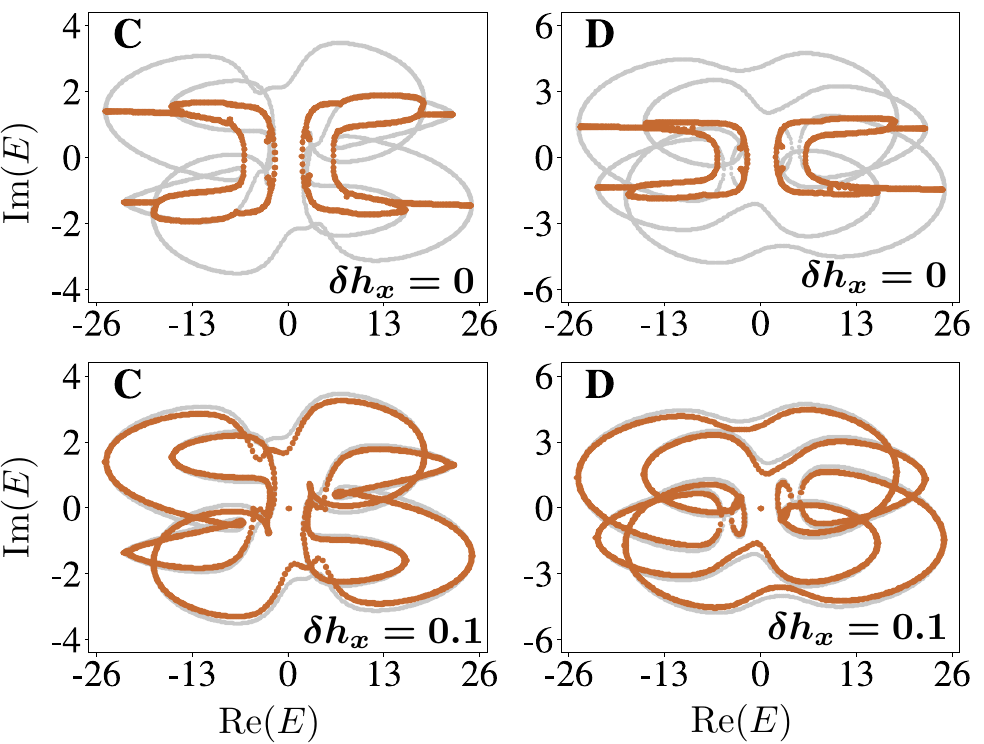}
    \caption{Energy spectra under PBC (gray curves) and OBC (brown dots) in the clean limit for  (top) $\delta h_x = 0$ and (bottom) $\delta h_x =0.1$. The left (right) column corresponds to point C (D) in Fig.~\ref{fig:phase_diagram}, with $g_2 = 3$ ($ g_2 = 6$), for which the winding number is $W=1$ ($W=2$). The adopted values of the remaining parameters are given by $4t_+=6$, $4t_-=10$, $g_+=3.0$, $g_-=3.5$, $\Gamma_+=3$, $\Gamma_-=3$, $\Delta_+=3$, $\Delta_-=-1.5$, $t_2 = 10$, and $N=500$.}
    \label{fig:spectra}
\end{figure}

To verify the consistency between the PBC winding numbers and the appearance of zero-energy modes under OBC, we present the PBC and OBC energy spectra for selected parameter sets in Fig.~\ref{fig:spectra}. These parameter sets, labeled as points C and D in Fig.~\ref{fig:phase_diagram}(c), correspond to phases with  different winding numbers. Similar spectra for the parameter sets labeled as points A and B are presented in Fig.~\ref{fig:Supp_B03_PD} in Appendix~\ref{sec:additional_results}.

In the absence of the perturbation, the OBC spectra in Fig.~\ref{fig:spectra} differ from the PBC spectra, revealing the NHSE. In contrast, the perturbation suppresses the NHSE, with the OBC spectra converging toward the corresponding PBC spectra. 
This suppression is also reflected in the bulk-state wave functions. As shown in Fig.~\ref{fig:OBC_varyN_avgIPR}(a) of Appendix~\ref{sec:additional_results}, the bulk states evolve from boundary-localized profiles to spatially extended profiles when the perturbation is turned on. We further compare the OBC and PBC spectra for different system sizes in the presence of the perturbation. As the system size increases, the bulk OBC spectra gradually approach the corresponding PBC spectra, indicating the suppression of the NHSE, as shown in Figs.~\ref{fig:OBC_varyN_avgIPR}(b)--\ref{fig:OBC_varyN_avgIPR}(d) of Appendix~\ref{sec:additional_results}. 
Accompanying the evolution of the bulk states, MZMs emerge in the OBC spectra, with one pair for parameter set C and two pairs for parameter set D, consistent with the winding number obtained from the PBC Hamiltonian. In the following section, we examine the properties of these MZMs.

\begin{figure}[t] 
    \centering
    \includegraphics[width=0.47\textwidth]{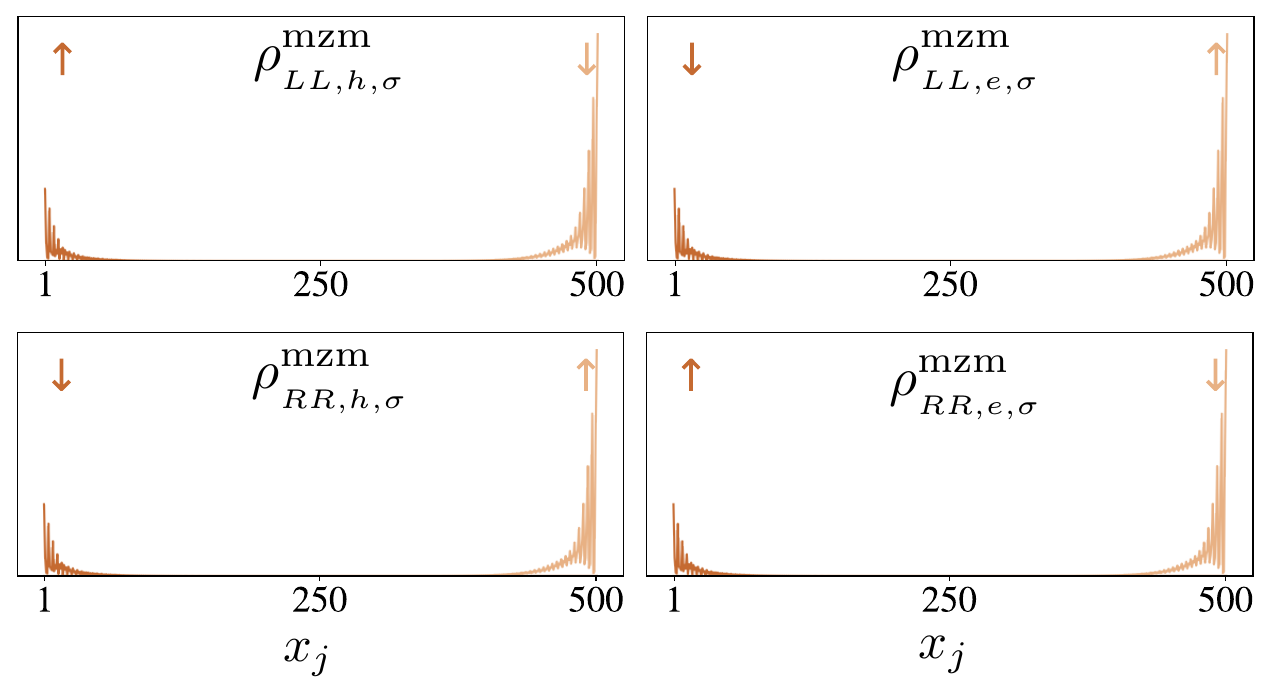}
    \caption{Spatial density profiles $\rho^{\rm mzm}_{RR (LL),\eta,\sigma} \equiv \rho^{E\approx 0}_{RR (LL),\eta,\sigma}$ for the particle ($e$)/hole ($h$) and for the up ($\uparrow$)/down ($\downarrow$) spin components of the right and left eigenstates of one MZM. Density-profile components related by symmetry are shown in the same color (see the main text). The adopted parameter values correspond to Point C in Fig.~\ref{fig:phase_diagram} with the lattice size $N=500$.
    }
    \label{fig:B03_C_densityprofile}
\end{figure}

\begin{figure*}[t]
    \centering
    \includegraphics[width=0.95\textwidth]{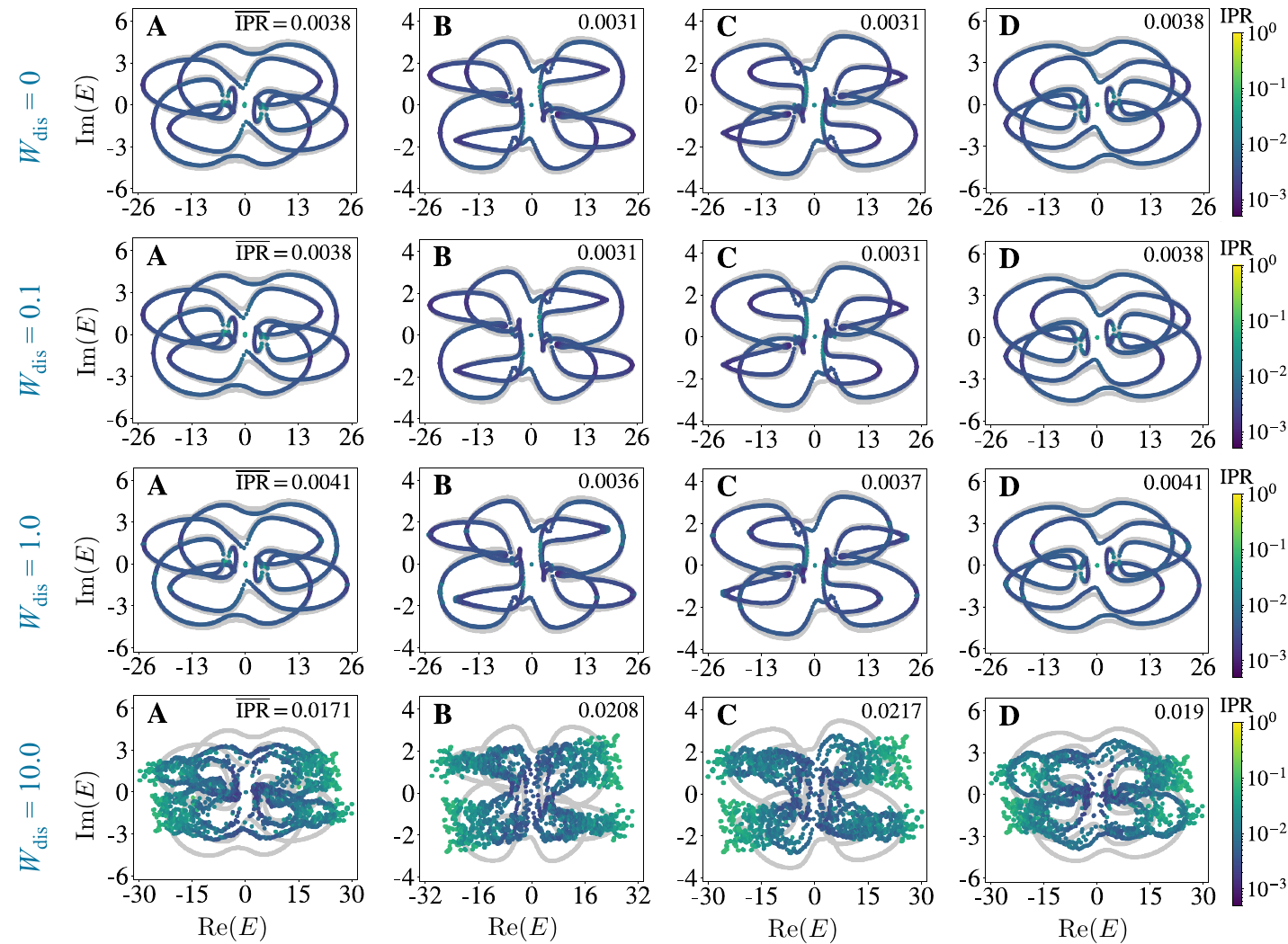}
    \caption{Evolution of OBC (colored) spectra for different disorder strengths $W_{\rm dis}$, compared with the PBC spectra (gray) in the clean limit. Each dot in the OBC spectra is color coded with the value of the IPR defined in Eq.~\eqref{eq:IPR}. From top to bottom, the rows correspond to $W_{\rm dis}\in\{0,\,0.1,\,1.0,\,10.0\}$, respectively, while the first to fourth columns correspond to parameter sets A–D indicated in Fig.~\ref{fig:phase_diagram}(c). All OBC spectra shown here are obtained from a single disorder realization. The values of  $\delta h_x=0.1$ and $N=500$ are adopted in all panels; the remaining parameters are provided in the caption of Fig.~\ref{fig:phase_diagram}. In each panel, the averaged IPR, defined in Eq.~\eqref{eq:IPR_avg}, is indicated.  
   }
    \label{fig:spectra_Dis}
\end{figure*}

\begin{figure}[t]
    \centering
    \includegraphics[width=0.48\textwidth]{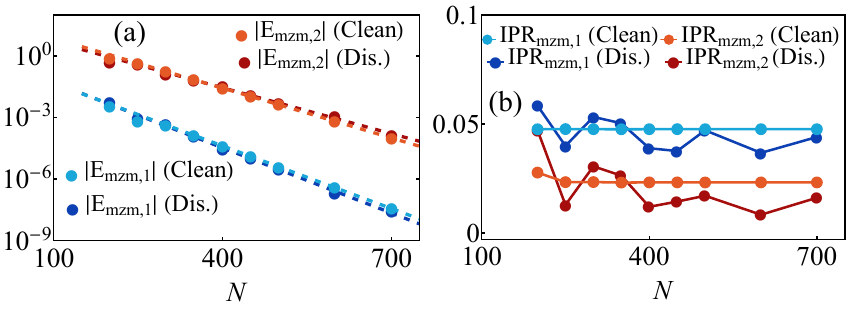}
    \caption{System-size ($N$) dependence of the near-zero modes for the model with sublattices [Eq.~\eqref{Eq:H_NHTSC}]. Panels (a) and (b) show the energy magnitudes $|E_{\rm mzm,1/2}|$ and the corresponding ${\rm IPR}_{\rm mzm,1/2}$, respectively. The dashed lines in panel (a) are exponential fits to $|E_{\rm mzm}|$. We use the parameters of point D in Fig.~\ref{fig:phase_diagram}(c), corresponding to a $W=2$ phase, with the remaining parameters given in the caption of Fig.~\ref{fig:phase_diagram}. Both clean and disordered cases are shown, with the disordered data denoted by ``Dis.'' in each panel. The disorder strength is $W_{\rm dis}=1.0$, and a single disorder realization is used.}
    \label{fig:finite-size scaling_NHTSC}
\end{figure}

\begin{figure}[t]
    \centering
    \includegraphics[width=0.48\textwidth]{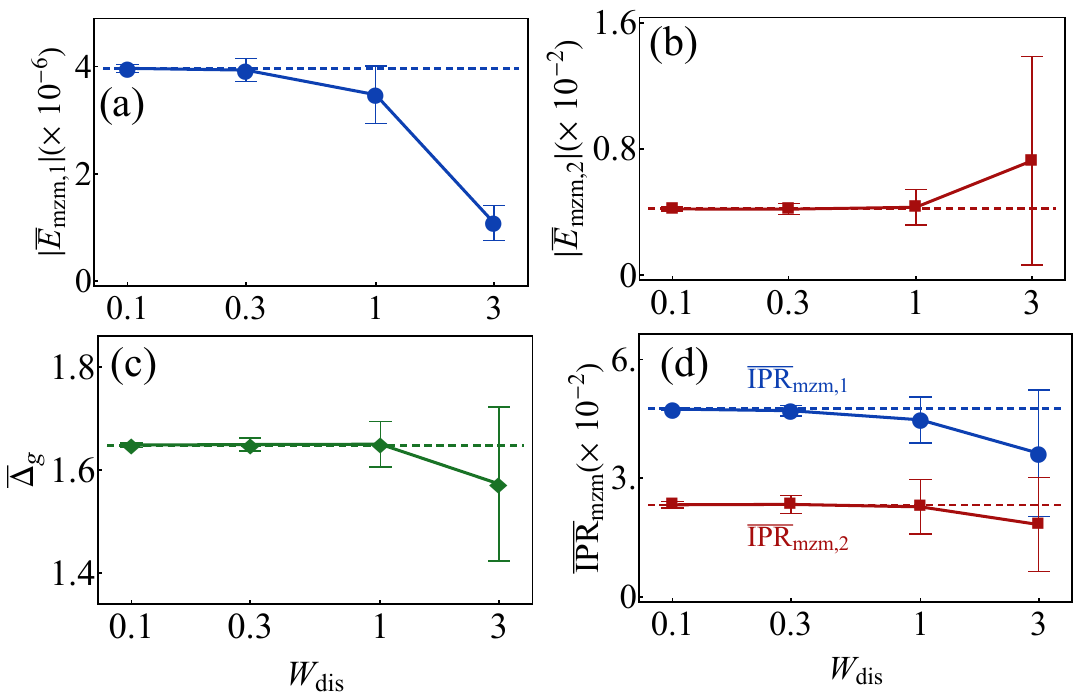}
    \caption{Disorder strength ($W_{\rm dis}$) dependence of disorder-averaged quantities, denoted by an overline for the model with sublattices [Eq.~\eqref{Eq:H_NHTSC}]. (a,b) Zero-mode energy magnitudes $|E_{\rm mzm,1/2}|$, (c) bulk gap $\Delta_g$, and (d) ${\rm IPR}_{\rm mzm}$. Error bars denote the standard deviation over five disorder realizations. The dashed lines indicate the corresponding  values at $W_{\rm dis}=0$. We fix $N=500$ and use the same model parameters as those in Fig.~\ref{fig:phase_diagram}(c).} 
    \label{fig:dis-avg}
\end{figure}

\subsection{Symmetry properties and stability of MZMs}
\label{eq:symm-enrich_stability}

In this section, we examine the properties of the MZMs identified in the OBC spectra. 
We first discuss their spatial profiles in the clean limit, analyze their system-size dependence, then perform a finite-size-scaling analysis, and finally discuss the stability of the MZMs in the presence of disorder.

Motivated by Ref.~\onlinecite{Chang:2026}, we consider the spatial dependence of the density profiles of the wave functions for the right ($|E\rangle$) and left ($\langle\!\langle E|$) eigenstates with the eigenenergy $E$. 
The site-resolved densities are defined as
\begin{align}
    \rho^E_{RR,\eta,\sigma} (x_j)
    &\equiv
    \langle E| x_j,\eta,\sigma \rangle
    \langle x_j,\eta,\sigma |E \rangle
    =
    |\psi_{\eta,\sigma}^E (x_j)|^2, \nn 
    \rho^E_{LL,\eta,\sigma} (x_j)
    &\equiv
    \langle\! \langle E| x_j,\eta,\sigma \rangle
    \langle x_j,\eta,\sigma |E \rangle\!\rangle
    =
    |\bar{\psi}_{\eta,\sigma}^E (x_j)|^2 .
    \label{eq:def_profile}
\end{align}
Here, $\psi_{\eta,\sigma}^E (x_j)$ and $\bar{\psi}_{\eta,\sigma}^E (x_j)$ denote the sector-resolved wave functions of the right and left eigenstates, respectively, obtained by projecting onto the basis state with spin $\sigma$ and particle-hole sector $\eta$ at site $x_j$. We focus on zero-energy modes and show a representative density profile of one of the MZMs in Fig.~\ref{fig:B03_C_densityprofile}, corresponding to the parameter set denoted by Point C in Fig.~\ref{fig:phase_diagram}(c). As expected for MZMs, both the right and left zero-mode wave functions are localized near the edges in all particle-hole and spin sectors.

Moreover, the density profiles in different particle-hole and spin sectors exhibit symmetry-enforced correlations. As shown in Fig.~\ref{fig:B03_C_densityprofile}, the density profile of the hole sector with spin $\sigma$ for the left eigenstate is identical to those of the hole sector with spin $-\sigma$ for the right eigenstate, the particle sector with spin $\sigma$ for the right eigenstate, and the particle sector with spin $-\sigma$ for the left eigenstate. The same relation holds upon interchanging the particle and hole sectors.
 
These correlations can be understood from the symmetries of the perturbed Hamiltonian. Using the symmetry operators listed in Eq.~\eqref{eq:Symm_pert_H_dhx}, one finds that the left and right eigenstates are connected by\footnote{In this footnote, we derive the first relation in Eq.~\eqref{eq:eigenstate_PHS_TRSd}. From
$U_{C_-}H_{\rm NHTSC}^T U_{C_-}^{-1}=-H_{\rm NHTSC}$, we obtain
$U_{C_-}H_{\rm NHTSC}^T=-H_{\rm NHTSC}U_{C_-}$, where
$H_{\rm NHTSC}\equiv H^{\rm obc}_{\rm NHTSC}+H_{\rm pt}$.
Acting on $|E\rangle\!\rangle^*$, which satisfies
$H_{\rm NHTSC}^T|E\rangle\!\rangle^*=E|E\rangle\!\rangle^*$, gives
$ 
U_{C_-}H_{\rm NHTSC}^T|E\rangle\!\rangle^*
=
E U_{C_-}|E\rangle\!\rangle^*
=
-H_{\rm NHTSC}U_{C_-}|E\rangle\!\rangle^* $.
Therefore, we have $
H_{\rm NHTSC}U_{C_-}|E\rangle\!\rangle^*
=
-E U_{C_-}|E\rangle\!\rangle^* $, 
which implies $U_{C_-}|E\rangle\!\rangle^*=|-E\rangle$. Equivalently, after replacing $E\to -E$, one obtains the first relation in Eq.~\eqref{eq:eigenstate_PHS_TRSd}. The remaining relations in Eq.~\eqref{eq:eigenstate_PHS_TRSd} can be derived in the same manner.
}
\begin{align}
    |E\rangle = U_{C_-} | -E \rangle \! \rangle^*,\quad
    |E\rangle = U_{C_+} | E \rangle \! \rangle^*,\quad
    |E\rangle = U_S |-E\rangle .
    \label{eq:eigenstate_PHS_TRSd}
\end{align}
For the MZMs, setting $E=0$ then directly relates the density profiles of the right and left zero-mode eigenstates in different particle-hole and spin sectors.

We first examine the relation between the opposite particle-hole sectors. Using the explicit form of $ U_{C_-}$, one can connect $\psi_{\eta ,\sigma}^E$ to  $  \big( \bar{\psi}_{-\eta,\sigma}^{-E}\big)^* $   on the same site for the same $\sigma$ and opposite $\eta$. 
Similarly, the operator $U_{C_+}$ leads to  $  \psi_{\eta ,\sigma}^E  \sim \eta \big( \bar{\psi}_{\eta,-\sigma}^{E} \big) ^*$, thus  relating  the density profiles with opposite spins but the same particle-hole sector. 
Finally, from the third relation in Eq.~\eqref{eq:eigenstate_PHS_TRSd}, we obtain the relation 
$      \psi_{\eta ,\sigma}^E  \sim  (-i\eta )\sigma  \psi_{-\eta ,-\sigma}^E $ and a similar relation for  $\bar{\psi}$. 
As a consequence, the internal symmetries lead to the correlations of the density profiles between different sectors in Fig.~\ref{fig:B03_C_densityprofile}. 

We now examine the stability of the MZMs against disorder. To this end, we include a disorder potential term,
\begin{align}
    \sum_{j,\sigma} (\delta \mu _{a,j} a^\dagger_{j,\sigma} a_{j,\sigma} + \delta \mu _{b,j} b^\dagger_{j,\sigma} b_{j,\sigma} ),
    \label{Eq:disorder_term}
\end{align}
where the onsite potential terms $\delta \mu _{a,j}$ and $\delta \mu _{b,j}$ are given by random real numbers  in the interval $  [-W_{\rm dis},W_{\rm dis}]$, with $W_{\rm dis}$ denoting the disorder strength.  
We check that disorder of this type preserves all the internal symmetries preserved in Eq.~\eqref{Eq:H_NHTSC} in the presence of perturbations; see Eq.~\eqref{eq:Symm_pert_H_dhx} and the details in Appendix~\ref{sec:App_symm}. 
Consequently, we expect that the zero modes remain robust against sufficiently weak disorder, provided that the bulk gap remains open.

To better characterize the localization behavior, we introduce the inverse participation ratio (IPR) of the right eigenstate for the energy $E$ and the average IPR,~\cite{Wang:2025}   given by
\begin{align}
    {\rm IPR}_E  & \equiv \sum_{x_j,\eta,\sigma} |\psi^E_{\eta,\sigma}(x_j)|^4 / | \langle E | E\rangle |^2  ,
    \label{eq:IPR} \\
    \overline{{\rm IPR}} & \equiv  \frac{1}{4N}\sum_E {\rm IPR}_E,  
    \label{eq:IPR_avg}
\end{align}
where a factor of $1/4$ in the second line arises from the degrees of freedom in the particle-hole and spin subspaces. The value of ${\rm IPR}_E$ lies in the range $[0,1]$, with a larger (smaller) value indicating a more localized (more extended) state. It therefore provides a convenient measure of localization. The averaged quantity $\overline{{\rm IPR}}$, obtained by averaging over eigenstates, characterizes the overall localization of the full spectrum and is used to compare data across different disorder strengths and system parameters. 

We present the OBC spectra with the perturbation in Fig.~\ref{fig:spectra_Dis} for
$W_{\rm dis}\in\{0,\,0.1,\,1.0,\,10\}$, showing the evolution from the clean limit to a strongly disordered regime. In Fig.~\ref{fig:spectra_Dis}, the OBC spectra are obtained from  a single disorder realization. For weak to intermediate disorder, namely $W_{\rm dis}=0.1$ and $1.0$ in the second and third rows, the overall OBC spectra remain nearly unchanged compared with the clean limit. The MZMs therefore remain robust in this regime. Consistently, both the state-resolved IPR and its spectral average change only weakly. This behavior is expected because the relevant clean-system energy scales for the cases shown in Fig.~\ref{fig:spectra_Dis} remain larger than $W_{\rm dis}=1.0$, allowing the topological character to persist. 

In contrast, when the disorder strength is further increased and exceeds the clean-limit bulk gap scale, the OBC spectra are strongly modified, as illustrated in the fourth row of Fig.~\ref{fig:spectra_Dis}. In this regime, the topological features are suppressed, and the states become strongly localized by disorder, as indicated by the large IPR values and the enhanced averaged IPR.

To proceed, we examine the system-size dependence of the magnitudes $|E_{\rm mzm}|$ of the complex energies for the near-zero modes and their associated IPRs, denoted by ${\rm IPR}_{\rm mzm}$. The results for both the clean case and a single disorder realization are shown in Fig.~\ref{fig:finite-size scaling_NHTSC}. As shown in Fig.~\ref{fig:finite-size scaling_NHTSC}(a), $|E_{\rm mzm}|$ for the two near-zero modes decreases exponentially with increasing system size in both the clean and disordered cases. This behavior is consistent with the exponentially suppressed overlap between Majorana wave functions localized at opposite ends of the chain, similar to the Hermitian systems.~\cite{Kitaev:2001,Albrecht:Nature_2016}
In the clean limit, both ${\rm IPR}_{\rm mzm}$ values in Fig.~\ref{fig:finite-size scaling_NHTSC}(b) rapidly saturate to nonzero values as the system size increases, indicating that the zero modes remain spatially localized in the thermodynamic limit. Disorder, on the other hand, tends to broaden the zero-mode wave functions. Nevertheless, the IPRs remain finite for the system sizes considered here, supporting the robustness of the localized zero modes. Taken together, the exponential suppression of $|E_{\rm mzm}|$ and the nearly size-independent finite ${\rm IPR}_{\rm mzm}$ values provide a more quantitative identification of the MZMs.

So far, the presented OBC spectra have been obtained from single disorder realizations. To quantify the robustness of the MZMs against disorder, we further perform disorder averaging over multiple realizations and examine the disorder-strength dependence of the zero-mode splittings, the OBC bulk gap, and the IPRs of the zero modes, as shown in Fig.~\ref{fig:dis-avg}. The zero-mode splittings remain much smaller than the bulk gap within the error bars, indicating that the near-zero modes remain well separated from the bulk spectrum. The disorder-averaged IPRs of the two pairs of MZMs also remain close to their clean-limit values, as shown in Fig.~\ref{fig:dis-avg}(d), suggesting that the zero modes remain spatially localized. The gradual decrease of the IPRs with increasing $W_{\rm dis}$ reflects the disorder-induced broadening of the zero-mode wave functions. These results indicate that the MZMs remain robust against sufficiently weak disorder.

\begin{figure}[t]
    \centering
     \includegraphics[width=0.48\textwidth]{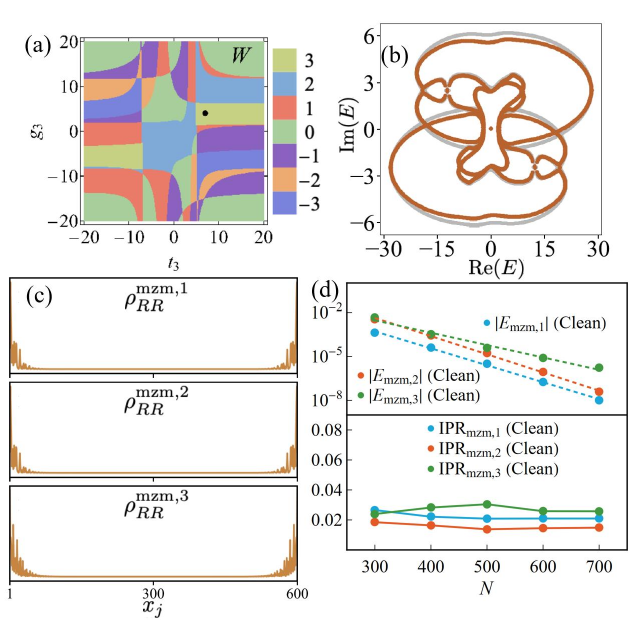}
    \caption{Representative results for the extended model, $H_{\rm NHTSC}+H_3$. (a) Topological phase diagram in the $(t_3,g_3)$ plane. (b) Energy spectra under PBC (gray) and OBC (brown) for $N=600$ at $(t_3,g_3)=(7.0,4.0)$ in the $W=3$ phase, marked by the dot in panel (a). (c) Right-eigenstate density profiles $\rho^{\rm mzm}_{RR}(x_j)$ for the three independent MZMs appearing in panel (b).  (d) Corresponding finite-size scaling of $|E_{\rm mzm}|$ and ${\rm IPR}_{\rm mzm}$. The remaining model parameters are fixed as $4t_+=4.0$, $4t_-=2.0$, $g_+=2.0$, $g_-=2.0$, $t_2=6.0$, $g_2=3.0$, $t_{3}^{\prime}=7.0$, $g_{3}^{\prime}=4.0$, $\Gamma_+=5.0$, $\Gamma_-=0.0$, $\Delta_+=2.5$, and $\Delta_-=0.0$. We use  $\delta h_x =0.1$ for panels (b)--(d).}
    \label{fig:NHTSC_sublattice_W3_phase_diagram}
\end{figure}

\subsection{Higher-winding phase with $W = 3$}
\label{sec:NHTSC_W_3_phase}

The model considered above includes longer-range hopping processes only up to second-nearest neighbors. Even higher winding numbers and additional pairs of MZMs can be realized by further extending the hopping range. To demonstrate this, we add third-nearest-neighbor hopping processes to Eq.~\eqref{Eq:H_NHTSC}. The full Hamiltonian is then given by $H_{\rm NHTSC}+H_3$, where
 \begin{align}
 H_3
 =
 & -\sum_{j,\sigma}\Big[
\left(t_3-\sigma\tfrac{g_3}{4}\right)
a_{j,\sigma}^{\dagger}b_{j+1,\sigma}
+
\left(t_3+\sigma\tfrac{g_3}{4}\right)
b_{j+1,\sigma}^{\dagger}a_{j,\sigma}
\nn
&
+
\left(t_3^{\prime}-\sigma\tfrac{g_3^{\prime}}{4}\right)
b_{j,\sigma}^{\dagger}a_{j+2,\sigma}
+
\left(t_3^{\prime}+\sigma\tfrac{g_3^{\prime}}{4}\right)
a_{j+2,\sigma}^{\dagger}b_{j,\sigma}
\Big].
\label{Eq:H3_NHTSC}
\end{align} 
Here, $t_3$ and $t_3^\prime$ are the third-nearest-neighbor hopping amplitudes across one and two unit cells, respectively, while $g_3$ and $g_3^\prime$ are the corresponding nonreciprocal hopping amplitudes. 

Since $H_3$ preserves all the internal symmetries of $H_{\rm NHTSC}$, the extended Hamiltonian remains in the same class characterized by a $\mathbb{Z}$ invariant. The winding number can therefore be constructed in the same way as in Eq.~\eqref{eq:W_nonHermi}. We evaluate this invariant and map out the topological phase diagram in the $(t_3,g_3)$ plane, as shown in Fig.~\ref{fig:NHTSC_sublattice_W3_phase_diagram}(a). As expected, including third-nearest-neighbor hopping processes gives rise to topological phases with $|W|=3$.

For a representative point in the $W=3$ phase, the OBC spectrum in Fig.~\ref{fig:NHTSC_sublattice_W3_phase_diagram}(b) and the corresponding density profiles\footnote{Figures~\ref{fig:NHTSC_sublattice_W3_phase_diagram}(c), \ref{Fig:GOS_W3_phase_diagram}(c), and \ref{fig:OBC_varyN_avgIPR}(a) show the total density profiles associated with the right eigenvectors, obtained by summing over the internal index,
\begin{align*}
\rho^{E}_{RR}(x_j)
=
\sum_{\eta,\sigma} \rho^{E}_{RR,\eta,\sigma}(x_j),
\end{align*}
where $\rho^{E}_{RR,\eta,\sigma}(x_j)$ is defined in Eq.~\eqref{eq:def_profile}. For the MZMs, we use $\rho^{\rm mzm}_{RR,\eta,\sigma}(x_j) \equiv \rho^{E\approx 0}_{RR,\eta,\sigma}(x_j)$.
} 
in Fig.~\ref{fig:NHTSC_sublattice_W3_phase_diagram}(c) show three pairs of MZMs localized near the system ends, consistent with the winding number. 
We remark that, for the model extended by even longer-range hopping, a larger system size $N$ is required to ensure that the three pairs of lowest-energy modes sufficiently approach zero energy in the OBC spectrum.
The finite-size scaling of $|E_{\rm mzm}|$ and ${\rm IPR}_{\rm mzm}$ in Fig.~\ref{fig:NHTSC_sublattice_W3_phase_diagram}(d) shows behavior similar to that in Fig.~\ref{fig:finite-size scaling_NHTSC}, further supporting the identification of these modes as MZMs.

 We have shown that the Schur-Cohn method provides an efficient way to compute winding numbers and determine the phase diagrams of 1D non-Hermitian topological superconductors characterized by point-gap topology, even when the PBC Hamiltonian takes the rather complicated form in Eq.~\eqref{Eq:H_pbc}. The same approach can also be extended to models with hopping terms beyond second-nearest neighbors.
 By including longer-range hopping terms, the same analysis can access regimes with even higher winding numbers, offering a systematic route to higher-winding   phases with multiple MZMs at each end.

\section{1D NHTSC model without sublattice}
\label{sec:NHTSC2}

\subsection{PBC Hamiltonian and symmetries}
\label{sec:Hamiltonian_symmetry_NHTSC2}

To demonstrate the broad applicability of the Schur--Cohn method, in this section we consider another example: a 1D NHTSC without a sublattice degree of freedom. The BdG form of the corresponding   Hamiltonian in momentum space is given by
\begin{eqnarray}
H_{\rm NHTSC,2}^{\rm pbc}(k)
&= &
\left[
-2t_1\cos(ka_0)
-2t_2\cos(2ka_0)
-\frac{i \Gamma_0}{2}
\right]\eta^z 
\nonumber
\\
&&
-\frac{i}{2}
\left[
g_1\sin(ka_0)
+
g_2\sin(2ka_0)
\right]\sigma^z
-\Delta_0 \,\eta^y \sigma^y , \nonumber \\ 
\label{Eq:H_NHTSC2_BdG}
\end{eqnarray} 
where $\eta^\mu$ and $\sigma^\mu$ are Pauli matrices acting on the particle-hole and spin degrees of freedom, respectively. The parameters have the same meanings as in the previous model, except that there is no distinction between intra-unit-cell and inter-unit-cell terms. The model in Eq.~\eqref{Eq:H_NHTSC2_BdG} can be viewed as an extension of the tight-binding Hamiltonian in Ref.~\citenum{Okuma:2019}, with the inclusion of longer-range hopping terms $t_2$ and $g_2$, as well as distinct onsite dissipation $\Gamma_0$ and nearest-neighbor nonreciprocal hopping $g_1$.
 
The symmetries and topological classification of Eq.~\eqref{Eq:H_NHTSC2_BdG} can be analyzed similarly to Sec.~\ref{Sec:NHTSC}. Similar to the previous example,  Eq.~\eqref{Eq:H_NHTSC2_BdG} possesses a nontrivial unitary symmetry, denoted here by $U_2=\eta^z\sigma^z$. The Hamiltonian can therefore be block diagonalized into sectors with $U_2$ eigenvalues $\pm 1$,
\begin{equation}
V_2^\dagger H_{\rm NHTSC,2}^{\rm pbc}(k) V_2
=
\begin{pmatrix}
H_{2,+}(k)&0\\
0&H_{2,-}(k)
\end{pmatrix},
\label{Eq:NHTSC2_block_Hpm_main}
\end{equation} 
where $V_2$ is the corresponding transformation matrix.
Each of the irreducible blocks belongs to class A with SLS, supporting a $\mathbb{Z}\oplus\mathbb{Z}$ topological invariant in the point-gap classification. 
The blocks $H_{2,\pm}(k)$ can be brought into the off-diagonal form,
\begin{equation}
 {H}_{2,\pm}(k)
=
\begin{pmatrix}
0 & h_{2,\pm}(k)\\
h^\prime_{2,\pm}(k)&0
\end{pmatrix} , 
\label{Eq:NHTSC2_Hpm_offdiag_main}
\end{equation}
and the corresponding winding numbers are defined as
\begin{subequations}
\label{eq:NHTSC2_w_pm_prime}
    \begin{eqnarray}
        w_{2,\pm}
    &=
    \int_0^{2\pi/a_0}
    \frac{dk}{2\pi i}\,
    \frac{d}{dk}
    \ln \bigl[h_{2,\pm}(k)\bigr] , \\
w^\prime_{2,\pm}
    &=
    \int_0^{2\pi/a_0}
    \frac{dk}{2\pi i}\,
    \frac{d}{dk}
    \ln \bigl[h^\prime_{2,\pm}(k)\bigr] .
    \end{eqnarray}
\end{subequations}
As before, we proceed by rewriting $h_{2,\pm}$ and $h_{2,\pm}^\prime$ as polynomials of $z = e^{i k a_0}$, and evaluating the corresponding winding numbers following the procedure in Appendices~\ref{Appendix:Schur-Cohn_detail} and~\ref{Appendix:deducing_phase_boundaries}.   

As in Sec.~\ref{Sec:NHTSC}, the two $U_2$ sectors are related by momentum inversion by TRS$^\dagger$. In particular, one finds 
\begin{equation}
h_{2,\pm } ^\prime (k) = h_{2,\mp} (-k),
\end{equation}
which implies $ 
  w_{2,\pm}^\prime = -\,w_{2,\mp}$.
Thus, the unperturbed model is characterized by a $\mathbb{Z}\oplus\mathbb{Z}$ invariant, which can be represented by $(w_{2,+},w_{2,-})$.

As in Sec.~\ref{Sec:NHTSC}, we also introduce a weak transverse magnetic field as a perturbation term to suppress the NHSE. The full PBC Hamiltonian is then given by
\begin{equation}
H_{\rm NHTSC,2}^{\rm pbc}(k)
+
\delta h_x\,\eta^z\sigma^x .
\end{equation}
The perturbation mixes the two spin sectors, thereby breaking the unitary symmetry and coupling the two previously independent blocks. 
As a result, the full Hamiltonian belongs to class D in the real AZ classification and class AI$^\dagger$ in the real AZ$^\dagger$ classification, together with SLS marked by $ S_+$. As before, once the NHSE is suppressed, the number of MZMs appearing in the OBC spectra is characterized by the winding number of the full system, $W_2 = w_{2,+} + w_{2,-}$.

\subsection{Topological phase diagram and energy spectra}
\label{sec:NHTSC2_winding_phase_diagram}
 
\begin{figure}[t]
    \centering
    \includegraphics[width=0.47\textwidth]{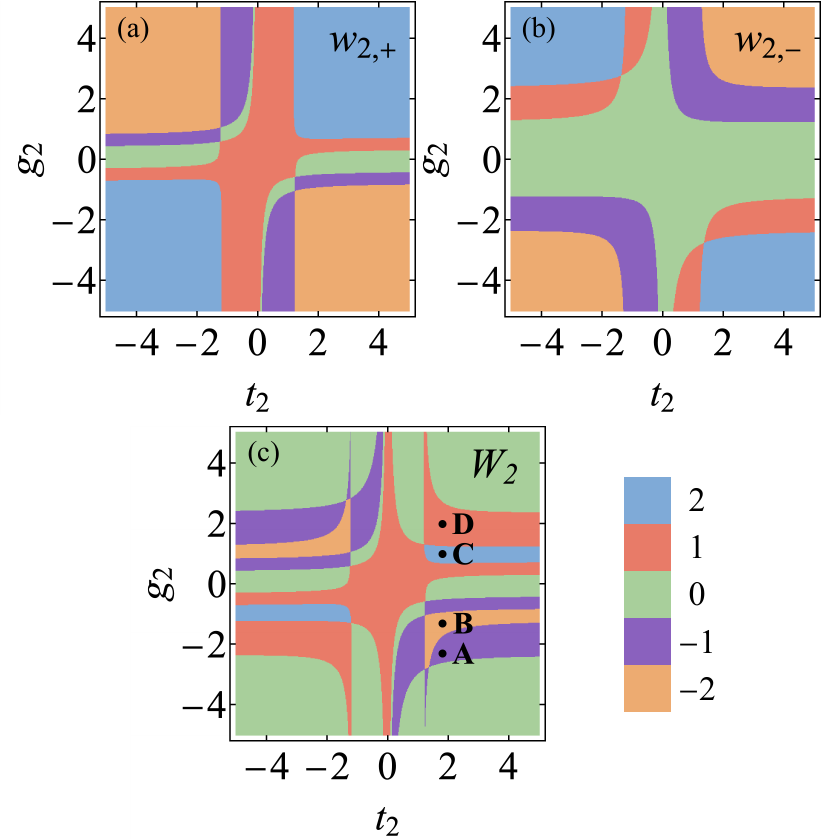}
    \caption{Phase diagrams based on the computed winding numbers from Eq.~\eqref{Eq:H_NHTSC2_BdG} in the $(t_2,g_2)$ plane. Panels (a) and (b) show $w_{2,+}$ and $w_{2,-}$ defined in Eq.~\eqref{eq:NHTSC2_w_pm_prime}, and Panel (c) shows their sum, $W_{2}$.
     The adopted values of the remaining parameters are given by $t_1=1.2$, $g_1= 0.8$,  $\Gamma_0=0.8$, and $\Delta_0=0.5$. 
     }
    \label{Fig:NHTSC2_phase_diagram}
\end{figure}

The topological phase diagrams obtained from the Schur-Cohn method are shown in Fig.~\ref{Fig:NHTSC2_phase_diagram}. The diagrams are plotted in the $(t_2,g_2)$ plane, with the remaining parameters fixed. As before, the inclusion of longer-range hopping terms generates phases with higher winding numbers over finite regions of parameter space. Consistently, as shown in Fig.~\ref{fig:PD_dhx}(b) in Appendix~\ref{sec:additional_results}, the phase diagram constructed in the presence of a small perturbation are almost unchanged compared with the unperturbed phase diagram shown in Fig.~\ref{Fig:NHTSC2_phase_diagram}(c), apart from small shifts in the phase boundaries.

\begin{figure}[t]
    \centering
    \includegraphics[width=0.47\textwidth]{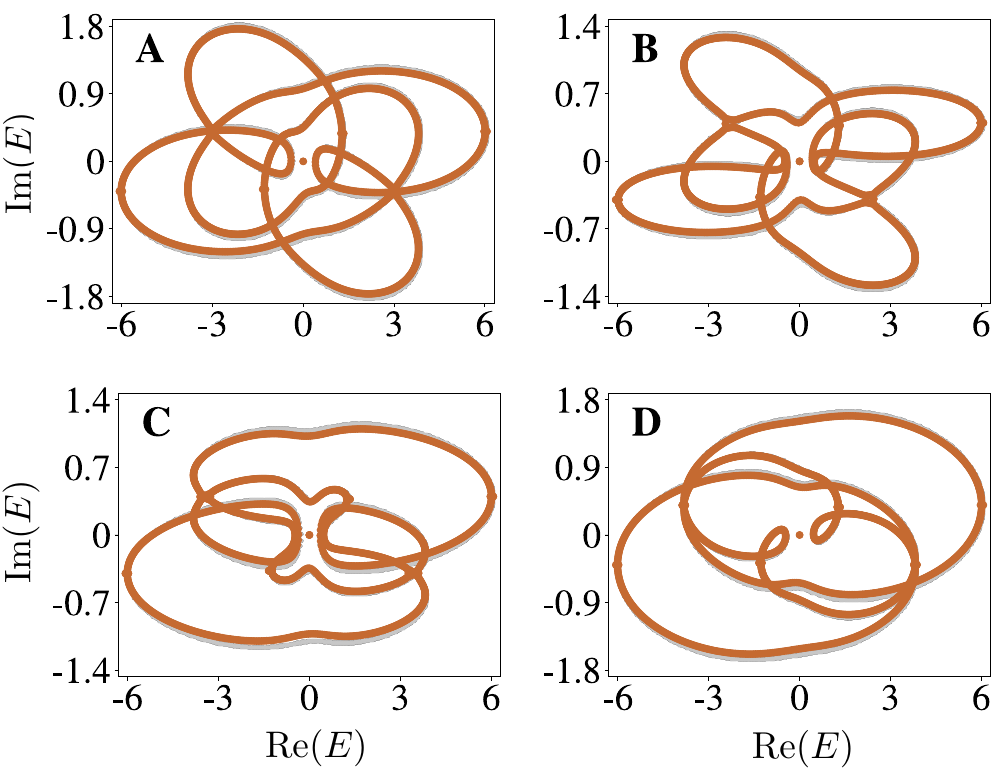} 
    \caption{Energy spectra under PBC (gray curves) and OBC (brown dots) of the model in Eq.~\eqref{Eq:H_NHTSC2_BdG} for $\delta h_x =0.05$ and $N=500$. The adopted value of $(t_2,g_2)$ in each panel is marked as A--D in Fig.~\ref{Fig:NHTSC2_phase_diagram}(c), corresponding to  $(t_2,g_2)$ with fixed $t_2 = 1.8$ and $g_2\in \{-2.3, \, -1.3, \, 1.0,\, 2.0 \}$. The remaining parameters used here are given in the caption of Fig.~\ref{Fig:NHTSC2_phase_diagram}. 
     }
    \label{fig:NHTSC2_spectra}
\end{figure}

\begin{figure}
    \centering
    \includegraphics[width=0.48\textwidth]{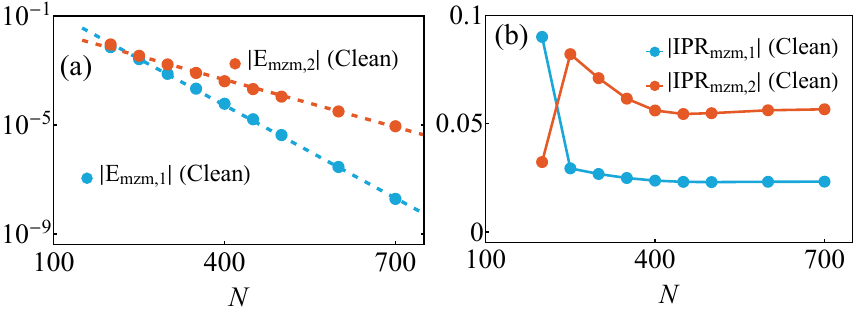}
    \caption{Similar plots to Fig.~\ref{fig:finite-size scaling_NHTSC}, but for the model $H_{{\rm NHTSC},2}$ without sublattice. We adopt the parameter set C given in Fig.~\ref{Fig:NHTSC2_phase_diagram}(c).}
    \label{fig:finite-size scaling_GOS}
\end{figure}

We also compute the OBC spectra for representative parameter sets in different phases with nonzero winding numbers. As before, in the absence of the perturbation, the NHSE generally emerges; these spectra are not shown for brevity. 
After introducing the perturbation, the sector-resolved NHSE is suppressed, as shown in Fig.~\ref{fig:NHTSC2_spectra}. For parameter sets in nontrivial phases, zero-energy boundary modes appear in the OBC spectra, with the number of MZM pairs consistent with the winding numbers shown in Fig.~\ref{Fig:NHTSC2_phase_diagram}.
A finite-size scaling analysis of the zero-energy modes presented in Fig.~\ref{fig:finite-size scaling_GOS} shows that their energy splittings decrease with increasing system size, while the corresponding IPRs converge to finite nonzero values.

\subsection{Higher-winding phase with $W_2 = 3$}
\label{sec:GOS_W_3_phase}

 To access phases with higher winding numbers, we follow the same strategy as in Sec.~\ref{sec:NHTSC_W_3_phase} and add third-nearest-neighbor hopping terms. The unperturbed PBC Hamiltonian in Eq.~\eqref{Eq:H_NHTSC2_BdG} is then extended to
\begin{align}
    H_{\rm NHTSC,2}^{\rm pbc}(k) -2t_3\cos(3ka_0) \eta_z -\frac{i}{2}g_3\sin(3ka_0) \sigma_z .
\label{Eq:H_NHTSC2_BdG_extended}
\end{align}
Since the extended Hamiltonian preserves the symmetries of $H_{\rm NHTSC,2}^{\rm pbc}(k)$, it remains in the same topological class, and the winding number can be defined as before.
 
\begin{figure}[t]
    \centering
    \includegraphics[width=0.48\textwidth]{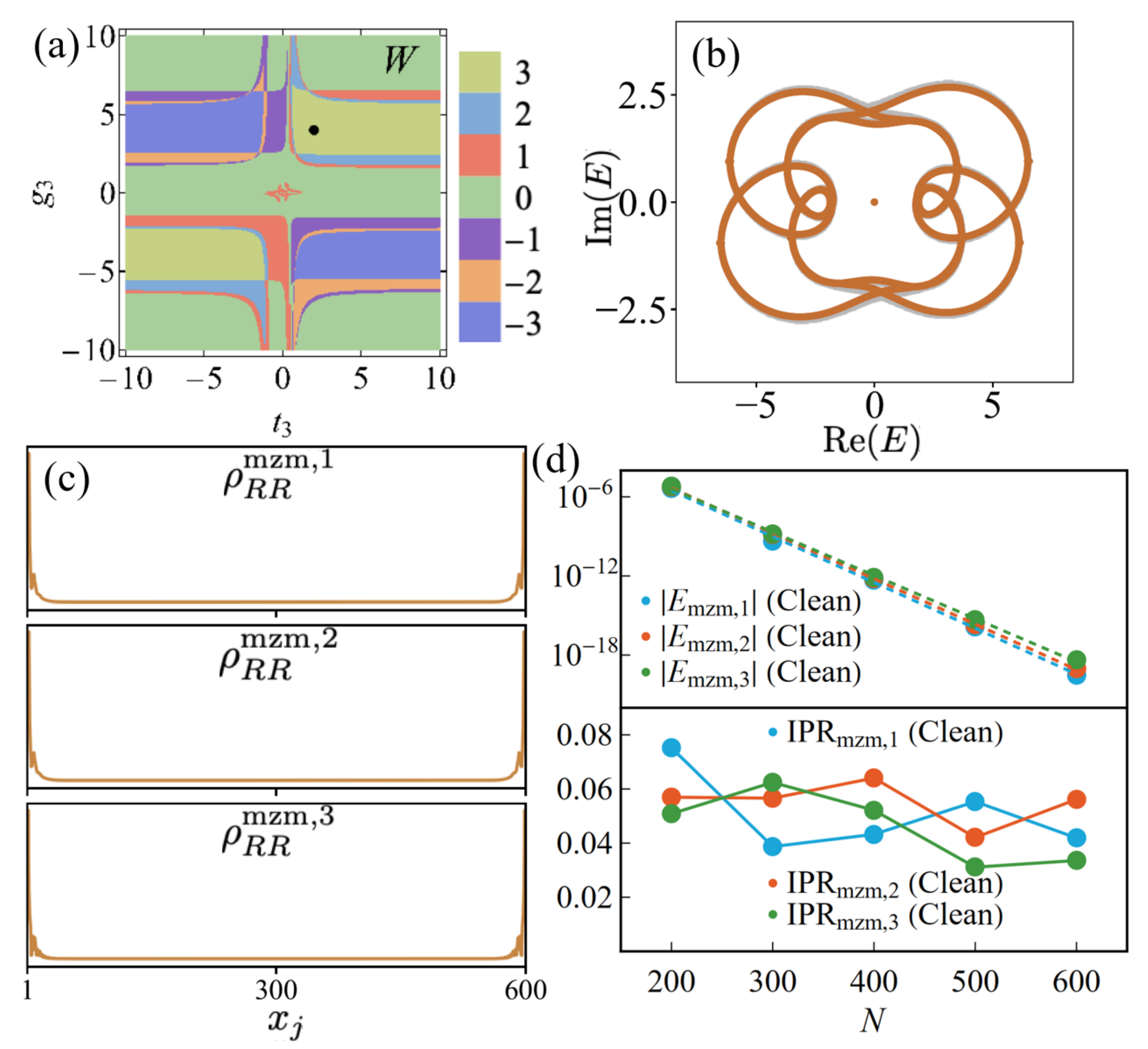}
    \caption{Similar plots to Fig.~\ref{fig:NHTSC_sublattice_W3_phase_diagram}, but for   Eq.~\eqref{Eq:H_NHTSC2_BdG_extended}. The system size is fixed at $N=600$ in panels (b) and (c).
    Panels (b)--(d) are evaluated using  $(t_3, \,g_3)= (2.0, \, 4.0)$ and $\delta h_x =0.05$, marked by the dot in  panel~(a). The remaining parameters  are $t_1=1.5$, $g_1=1.0$, $t_2=0.1$, $g_2=0.1$, $\Gamma_0=2.0$, and $\Delta_0=2.0$.}
    \label{Fig:GOS_W3_phase_diagram}
\end{figure}

Figure~\ref{Fig:GOS_W3_phase_diagram}(a) shows the corresponding phase diagram in the $(t_3,g_3)$ plane. The additional $t_3$ and $g_3$ terms generate a $|W_2|=3$ phase over a finite parameter region. As shown in Figs.~\ref{Fig:GOS_W3_phase_diagram}(b) and (c), once the NHSE is suppressed, three pairs of MZMs appear at the system boundaries, consistent with the winding number. This identification is further supported by the finite-size scaling shown in Fig.~\ref{Fig:GOS_W3_phase_diagram}(d).

\section{Discussion and conclusions}
\label{Sec:Discussion}

In this work, we adopted the Schur-Cohn method to compute winding numbers and construct topological phase diagrams for 1D non-Hermitian  systems. The central idea is to convert the winding-number calculation into a problem of counting the zeros of a characteristic polynomial, expressed directly in terms of the system parameters. This avoids both the explicit evaluation of the winding integral and the direct solution of the PBC gap-closing conditions. For models with large internal degrees of freedom or longer-range hopping terms, such direct approaches quickly become impractical over broad parameter space. The Schur-Cohn method therefore provides a systematic and efficient alternative.

The usefulness of this approach is demonstrated in both example models studied here. Starting from the characteristic polynomials, we obtained the winding numbers and phase boundaries without explicitly solving the PBC energy spectrum or deriving the gap-closing conditions from it.
This coefficient-based formulation is particularly advantageous for longer-range models with large internal degrees of freedom, such as particle-hole, sublattice and spin, where the relevant polynomials become higher order and closed-form expressions for their roots are generally unavailable. It also makes it possible to identify phases with even higher winding numbers in a controlled way.

In non-Hermitian systems, the energy spectrum is generally complex, and topological phases can be characterized by point-gap structures and spectral winding numbers that have no direct analogue in Hermitian systems. The polynomial method used here provides an analytic way to track how the zeros of the relevant complex polynomial move as non-Hermitian parameters are varied. In particular, non-Hermitian terms modify the complex coefficients of the polynomial and may induce crossings of the unit circle $|z|=1$ that are absent in the Hermitian limit. These crossings mark PBC point-gap-closing conditions and can separate distinct topological regions. Thus, the method helps identify additional non-Hermitian topological phases, while their physical origin comes from the non-Hermitian structure of the Hamiltonian itself.

We also clarify the origin of higher-winding phases. In 1D tight-binding models, the winding number is determined by how the relevant complex function, such as a determinant, winds as $k$ varies over the Brillouin zone. Higher winding numbers therefore typically require sufficiently rich momentum dependence, for example from longer-range hopping terms that generate higher harmonics such as $\cos(2ka_0)$, or equivalently higher powers of $z=e^{ika_0}$ in the complex polynomial. Nonreciprocal and dissipation terms can further shift the polynomial zeros relative to the Hermitian limit and thereby modify the resulting winding structure. 
As demonstrated in this work, for systems characterized by $\mathbb{Z}$ topology, the inclusion of longer-range hopping processes allows topological phases with $|W|>1$ to appear. These phases host multiple MZMs at each end and therefore go beyond the parameter regimes explored in Refs.~\onlinecite{Okuma:2019,Chang:2026}, where the relevant phases were characterized by $|W|\le 1$ within the considered parameter space. In this sense, the present analysis extends the non-Hermitian topological superconductor framework to higher-winding sectors and provides a practical analytic route for mapping their phase diagrams.

We also clarified the role of the weak onsite transverse perturbation in suppressing the NHSE. Reference~\onlinecite{Okuma:2019} identified the conditions under which an infinitesimal perturbation suppresses the NHSE. Here, we provide an explicit derivation of this condition within our formulation and show how the winding numbers before and after the perturbation are related. This establishes a direct connection between the winding numbers associated with skin modes in the unperturbed system and those associated with MZMs after the NHSE is suppressed. 
In other words, although the conventional bulk-boundary correspondence is generally not directly applicable to non-Hermitian systems with the NHSE, it can be recovered once the NHSE is sufficiently suppressed. This makes a direct comparison between PBC winding numbers and OBC boundary modes meaningful in the perturbed system.

More broadly, the coefficient-based approach adopted here provides a practical route to higher-winding point-gap topology in non-Hermitian systems with enlarged internal structures, where direct spectral analysis becomes impractical. This provides a systematic framework for mapping phase boundaries and identifying the corresponding boundary zero modes in higher-winding non-Hermitian topological superconductors.

\begin{acknowledgments}
We thank C.-K.~Chang, N.~Okuma, K.~Saito, and K.~Wakabayashi for interesting discussions. We thank H.-C.~Wang for inventing the symbol of $\intra$, which is adopted to label the intracell nearest-neighbor terms.
This work was financially supported by the National Science and Technology Council (NSTC), Taiwan (Grant No.~NSTC-114-2112-M-001-057, Grant No.~NSTC-114-2811-M-001-051, and Grant No.~NSTC-115-2112-M-001-024) and Academia Sinica (AS), Taiwan  (Grant No.~AS-iMATE-114-12), and JSPS KAKENHI (Grant Nos.~JP22H05118, JP26K00629, and JP26H013).  
We acknowledge the technical support from the Academia Sinica Grid Computing Center (ASGC), Taiwan, through Grant No.~AS-CFII-112-103.

\end{acknowledgments}

\section*{Author Declarations}
\subsection*{Conflict of Interest Statement}
The authors declare no conflict of interest.
\subsection*{Author Contributions}
Yung-Yeh Chang and Xiang-Yu Li contributed equally to this work.
\textbf{Yung-Yeh Chang}:  Data curation (equal); Formal analysis (equal); Investigation (equal); Validation (equal); Supervision (supporting); Writing - original draft (equal); Writing - review and editing (equal).
\textbf{Xiang-Yu Li}:  Data curation (equal); Formal analysis (equal); Investigation (equal); Validation (equal);  Writing - original draft (equal); Writing - review and editing (equal).
\textbf{Ken Shiozaki}: Validation (equal); Funding Acquisition (lead); Writing - review and editing (equal).
\textbf{Chen-Hsuan Hsu}: Conceptualization (lead); Investigation (equal); Validation (equal); Funding Acquisition (lead); Supervision (lead); Writing - original draft (equal); Writing - review and editing (equal).

\section*{Data Availability Statement}
The data that support the findings of this study are openly available in Zenodo at Ref.~\citenum{zenodo_data}.

\appendix

\section{Details about the Schur-Cohn method}
\label{Appendix:Schur-Cohn_detail}
 
In this section, we provide the details about evaluating the winding number based on the Schur-Cohn method, including the coefficient-based procedure and its mathematical justification. 
As discussed in Sec.~\ref{Sec:Winding_Evaluation}, the task is to  count the number of zeros of a complex polynomial inside the unit circle. Here we discuss the procedure to determine this number directly from the polynomial coefficients without explicitly solving the corresponding polynomial equation.~\cite{Schur:1917,Cohn:1922,Henrici:1974,Rahman:2002,Stoica:1992}

We now consider an arbitrary complex polynomial of degree \(n\)  [that is, $P_{2d}(z)$ defined in Sec.~\ref{Sec:Winding_Evaluation}], 
\begin{equation}
p_n(z)=c_0+c_1 z+\cdots +c_n z^n, \qquad c_n \neq 0,
\label{Eq:p(z)_app}
\end{equation}
and introduce its reciprocal counterpart as 
\begin{equation}
\overline{p}_n(z)=z^n [p_n(1/z^*)]^*, 
\end{equation}
where $z^*$ denotes the complex conjugate of $z$. 
The two polynomials satisfy \(|\overline{p}_n(z)|=|p_n(z)|\) on the unit circle \(|z|=1\). In addition, applying the reciprocal operation twice recovers the original polynomial, i.e. $\overline{\overline{p}}(z) = p(z)$. 
We further denote the number of zeros of \(p_n(z)\) inside \(|z|<1\) by \(\mathbb{N} [p_n(z)]\), counted with multiplicity. As a consequence, the zeros of \(\overline{p}_n(z)\) inside the unit circle correspond to the zeros of \(p_n(z)\) outside the unit circle.

The Schur-Cohn method is based on the iterative transformation, $T_{\rm sc}$, which acts on the polynomial and gives 
\begin{equation}
T_{\rm sc} [p_n(z)]=c^*_0\,p_n(z)-c_n\,\overline{p}_n(z),
\end{equation}
which reduces the degree of the polynomial. Repeating this transformation for $\chi$ times generates a sequence of polynomials \(T_{\rm sc}^\chi [p_n(z)]\) and a corresponding sequence of  real quantities,  
\begin{equation} 
\gamma_\chi \equiv T_{\rm sc}^\chi [p_n(z)] \Big|_{|z| = 0}, \qquad \chi \in \{ 1,2,\dots,n \}.
\end{equation}
The sequence \(\{\gamma_\chi\}\)   encodes the distribution of zeros of \(p_n(z)\) relative to the unit circle.

In the following, we proceed with the case $\gamma_\chi\neq 0$ for all $\chi$. If $\gamma_\chi=0$ for some $\chi$, the above recursion procedure becomes inapplicable. Such cases signal possible changes in the winding number, or equivalently candidate gap-closing conditions, and will be discussed in Appendix~\ref{Appendix:deducing_phase_boundaries}. 
We further assume that none of the intermediate polynomials \(T^\chi_{\rm sc}[p_n(z)]\) has zeros on the unit circle  \(|z|=1\), such that Rouché's theorem~\cite{Ahlfors:1979,Henrici:1974}
can be applied on the unit circle. 
 
To proceed, we consider the  polynomial  after $\chi$ iterations, 
\begin{equation}
T^\chi_{\rm sc}[p_n(z)]
=
c^{(\chi)}_0+c^{(\chi)}_1z+\cdots+c^{(\chi)}_{n-\chi}z^{n-\chi}, 
\end{equation}
and also its next iteration, 
\begin{equation}
\label{Eq:chi+1}
T^{\chi+1}_{\rm sc}[p_n(z)]
=
\left(c^{(\chi)}_0\right)^*
T^\chi_{\rm sc}[p_n(z)]
-
c^{(\chi)}_{n-\chi}
\overline{T^\chi_{\rm sc} [ p_n (z) ] } . 
\end{equation}
On the unit circle, the reciprocal polynomial satisfies
\begin{equation}
\left|
\overline{T^\chi_{\rm sc} [ p_n (z) ] }
\right|_{|z| = 1}
=
\Big|
T^\chi_{\rm sc}[p_n(z)]
\Big|_{|z| = 1}.
\end{equation}
Therefore, on the unit circle, the relative magnitudes of the two terms on the right-hand side of Eq.~\eqref{Eq:chi+1}  are determined solely by comparing the magnitudes of $c^{(\chi)}_0$ and $c^{(\chi)}_{n-\chi}$. 
Equivalently, this comparison is encoded in the real quantity in the $(\chi+1)$th iteration, 
\begin{equation}
\gamma_{\chi+1}
=
\left|
c^{(\chi)}_0
\right|^2
-
\left|
c^{(\chi)}_{n-\chi}
\right|^2 .
\end{equation}

Therefore, the numerical procedure involves  checking  the sign of $\gamma_{\chi+1}$. 
If \(\gamma_{\chi+1}>0\), we obtain 
\begin{equation}
\left|
\left(c^{(\chi)}_0\right)^*
T^\chi_{\rm sc}[p_n(z)]
\right|\Big|_{|z| = 1}
>
\left|
c^{(\chi)}_{n-\chi}
\overline{T^\chi_{\rm sc} [ p_n (z) ] }
\right|\Big|_{|z| = 1}, 
\end{equation}
meaning that the first term is the dominant term. 
Applying Rouch{\'e}'s theorem,~\footnote{
The theorem states that if two analytic functions \(f_1(z)\) and \(f_2(z)\) satisfy $ 
|f_2(z)|<|f_1(z)| $ 
on a closed contour, then \(f_1(z)\) and \(f_1(z)+f_2(z)\) have the same number of zeros inside the contour.
}
we  find that  $T^{\chi+1}_{\rm sc}[p_n(z)]$  has the same number of zeros inside the unit circle  as \(T^\chi_{\rm sc}[p_n(z)]\):
\begin{equation}
\mathbb{N}
\!\left( 
T^{\chi+1}_{\rm sc} [p_n (z)]
\right)
=
\mathbb{N}
\!\left(
T^\chi_{\rm sc} [p_n (z)]
\right), \quad {\rm for} \;  \gamma_{\chi+1}>0. 
\end{equation}

In contrast, if $\gamma_{\chi+1}<0$, the second term is dominant. 
Using the reciprocal-zero correspondence, the number of zeros of the reciprocal polynomial inside the unit disk is equal to the number of zeros of \(T^\chi_{\rm sc}[p_n(z)]\) outside the unit disk. 
Since \(T^\chi_{\rm sc}[p_n(z)]\) has degree $ (n-\chi)$, we obtain
\begin{align}
\mathbb{N}
\!\left( 
T^{\chi+1}_{\rm sc} [p_n (z)]
\right)
 & =
\mathbb{N}
\!\left( 
\overline{T^\chi_{\rm sc} [ p_n (z) ] }
\right)
\nn
& =
(n-\chi)
-
\mathbb{N}
\!\left(
T^\chi_{\rm sc} [p_n (z)] 
\right) , \quad {\rm for} \;  \gamma_{\chi+1} < 0. 
\label{eq:N_chi_iteration}
\end{align}

We summarize these relations, with a shift in the dummy index, to obtain  
\begin{equation}
\mathbb{N}
\!\left[
T^{\chi}_{\rm sc} p_n  (z)
\right]
=
\begin{cases}
\mathbb{N}
\!\left[
 T^{\chi-1}_{\rm sc} p_n (z)
\right],
&
\gamma_\chi>0,
\\[6pt]
n-\chi + 1  
-
\mathbb{N}
\!\left[
 T^{\chi-1}_{\rm sc}p_n  (z)
\right],
&
\gamma_\chi<0.
\end{cases}
\end{equation}
Therefore, provided that $\gamma_{\chi} \neq 0$ for any $\chi$, the operation \(T^n_{\rm sc}[p_n(z)]\) leads to a nonzero constant after \(n\) iterations, giving $\mathbb{N}\!\left[(T^n_{\rm sc}p_n) (z)\right]=0$. 

To determine the final value of $\mathbb{N}[p_n(z)]$, we collect the terms with $\gamma_\chi<0$. Namely, we have a set, for $\chi\in\{\chi_1,\chi_2,\cdots,\chi_m\}$ such that $\gamma_\chi<0$, ordered as $\chi_1<\chi_2<\cdots<\chi_m$. By iterating the recursion backward, we obtain
\begin{equation}
\mathbb{N}[p_n (z)]
=
\sum_{\ell =1}^{m}
(-1)^{\ell-1}
\left(
n+1-\chi_\ell
\right), \;\; {\rm with } \;\; \gamma_{\chi_1} , \cdots, \gamma_{\chi_m} <0 
\label{Eq:Schur_Cohn_closed_formula}
\end{equation}
This is equivalent to Theorem~6.8c in Ref.~\citenum{Henrici:1974}. 
It shows that the number of zeros inside the unit circle can be obtained entirely from the sign pattern of the sequence $\{\gamma_\chi\}$, and hence from the coefficients of the polynomial. Thus, the winding number can be evaluated without explicitly solving the polynomial equation, determining the gap-closing conditions, or solving the PBC eigenvalue problem.
In Secs.~\ref{Sec:NHTSC}-\ref{sec:NHTSC2}, we apply this method to numerically evaluate the winding numbers for the specific models and verify the results by comparison with spectra obtained independently from the OBC Hamiltonians.

As a remark, the standard Schur-Cohn procedure presented here assumes that the intermediate quantities $\gamma_\chi$ are nonzero. The above procedure and Eq.~\eqref{Eq:Schur_Cohn_closed_formula} are therefore not directly applicable when $\gamma_\chi=0$, since the inequality required by Rouché's theorem is no longer valid. In physical models, such cases are related to possible point-gap closing conditions and are therefore analyzed together with the conditions for phase boundaries. These conditions can be handled by a complementary method, as demonstrated in Appendix~\ref{Appendix:deducing_phase_boundaries}.

\section{Analytic derivation of phase boundaries through a real-polynomial method}
\label{Appendix:deducing_phase_boundaries}

In this section, we discuss a real-polynomial method~\cite{Ahlfors:1979,Lang:2013,Cox:1997,Gelfand:1994} to analytically derive the conditions for topological phase boundaries, complementary to Eq.~\eqref{Eq:phase_boundary_half-analytic}. While this approach is more involved, it is more direct, as it imposes the unit-circle condition explicitly and thereby avoids extraneous solutions from off-unit-circle reciprocal root pairs.

As discussed in the main text, a PBC gap closes when a zero of the original degree-\(n\) polynomial \(p_n(z)\) [that is, $P_{2d}(z)$ introduced in Sec.~\ref{Sec:Winding_Evaluation}] lies on the unit circle. The topological phase boundary is therefore determined by
\begin{equation}
p_n(z_0)=0,\qquad |z_0|=1 .
\label{Eq:phase_boundary_unit_circle_condition}
\end{equation}
We now describe how this condition can be converted into an explicit expression in terms of the model parameters. To parametrize the unit circle, we use the Cayley transformation
\begin{equation}
z=\frac{1+i x}{1-i x}, \qquad x\in {R},
\label{Eq:Cayley_transform_phase_boundary}
\end{equation}
which satisfies $|z|=1$ and maps the real axis to the unit circle. Conversely, every point on the unit circle except $z=-1$ can be represented by Eq.~\eqref{Eq:Cayley_transform_phase_boundary}; 
the missing point corresponds to $x=\infty$ and can be checked separately by imposing $p_n(-1)=0$.

To proceed, we define the Cayley-transformed polynomial
\begin{equation}
Q_n(x)
=
(1-i x)^n
p_n\!\left(
\frac{1+i x}{1-i x}
\right).
\label{Eq:Cayley_transformed_polynomial}
\end{equation}
Since the prefactor \((1-i x)^n\) is nonzero for finite  \(x\), the condition \(Q_n(x)=0\) is equivalent to \(p_n(z)=0\) with \(z\) given by Eq.~\eqref{Eq:Cayley_transform_phase_boundary}. 
Therefore, for finite \(x\), a zero of \(p_n(z)\) on the unit circle is equivalent to a real zero of \(Q_n(x)\).

Using Eq.~\eqref{Eq:p(z)_app}, the polynomial \(Q_n(x)\) can be written as
\begin{equation}
Q_n(x)
=
\sum_{\ell=0}^{n}
c_\ell
(1+i x)^\ell
(1-i x)^{n-\ell} \equiv \sum_{\gamma=0}^{n} d_\gamma x^\gamma ,  
\label{Eq:Cayley_transformed_polynomial_expanded}
\end{equation} 
with $ \gamma \in \{ 0,1,\dots,n\}$ and the complex coefficient,
\begin{equation}
d_\gamma
=
i^\gamma
\sum_{\ell=0}^{n}
c_\ell
\sum_{\alpha+\beta= \gamma  }
\, 
\sum_{ 0\leq \alpha\leq \ell}
\,
\sum_{ 0\leq \beta\leq n-\ell}
(-1)^\beta
\binom{\ell}{\alpha}
\binom{n-\ell}{\beta}, 
\label{Eq:Cayley_coefficients_general}
\end{equation}
obtained by explicitly expanding the second expression in Eq.~\eqref{Eq:Cayley_transformed_polynomial_expanded} and applying  the binomial theorem. 
The above relation shows that  \(d_\gamma\) can be obtained through algebraic combinations of the original coefficients \(c_\ell\).

Since \(Q_n(x)\) is generally complex-valued, we separate its real ($R$) and imaginary ($I$) parts by defining
\begin{equation}
Q_n(x)=Q_{n}^R(x) + iQ_{n}^I(x),
\end{equation}
with two real polynomials, \(Q_{n}^R(x)\) and \(Q_{n}^I(x)\).
This implies that the equation \(Q_n(x_0)=0\) is equivalent to
\begin{equation}
Q_{n}^R(x_0)=0,
\qquad
Q_{n}^I(x_0)=0.
\label{Eq:real_imag_common_root}
\end{equation}
Thus, the unit-circle-zero condition is converted into the condition that $Q_n^R$ and $Q_n^I$ have a common real root.

Instead of solving Eq.~\eqref{Eq:real_imag_common_root} explicitly, we first make use of the resultant of $Q_{n}^R$ and $Q_{n}^I$. 
In practice, this resultant is evaluated as the determinant of the Sylvester matrix constructed from the coefficients of $Q_n^R$ and $Q_n^I$,~\cite{Ahlfors:1979,Lang:2013,Cox:1997,Gelfand:1994}
\begin{equation}
\operatorname{Res}_x [ Q_{n}^R(x),Q_{n}^I(x) ]
=
\det \operatorname{Syl}[ Q_{n}^R(x),Q_{n}^I(x) ].
\label{Eq:resultant_as_Sylvester_det}
\end{equation}
The key property we use is that a vanishing determinant of the Sylvester matrix implies that $Q_{n}^R$ and $Q_{n}^I$ are not coprime and therefore have a common root. Thus, the algebraic condition for the existence of a common root, $ 
\operatorname{Res}_x[ Q_{n}^R(x),Q_{n}^I(x) ] =0 $, can be equivalently written as
\begin{equation}
\det \left| \operatorname{Syl}[ Q_{n}^R(x),Q_{n}^I(x) ] \right| 
=
0 .
\label{Eq:resultant_phase_boundary}
\end{equation}
Since the coefficients of $Q_{n}^R$ and $Q_{n}^I$ are algebraic combinations of the coefficients $c_\ell$ of $p_n(z)$, the final expression is an algebraic equation written only in terms of $c_\ell$, their complex conjugates, and hence the model parameters. In consequence, the condition for the topological phase boundaries is obtained directly from the coefficients of the polynomial, without explicitly solving any polynomial equations. 

We emphasize that Eq.~\eqref{Eq:resultant_phase_boundary} gives only a necessary condition for a phase boundary. Although the vanishing resultant in Eq.~\eqref{Eq:resultant_phase_boundary} guarantees that $Q_n^R(x)$ and $Q_n^I(x)$ have a common root, this common root need not be real. Such a solution does not correspond to a physical point on the unit circle of $z$. Therefore, after obtaining candidate phase boundaries from Eq.~\eqref{Eq:resultant_phase_boundary}, we retain only those for which $Q_n^R(x)$ and $Q_n^I(x)$ have at least one real common root. This additional real-root condition removes extraneous solutions produced by common complex roots.

In the calculations presented in this work, the above procedure is used to generate candidate phase-boundary curves. We then apply the real-root criterion and also include the separate condition associated with $z=-1$. 
We have checked that this method reproduces the conditions for the models investigated in this work.

\section{Details about the 1D NHTSC model with sublattice}
\label{Appendix:Details_NHTSC}

In this section, we present the details about the 1D NHTSC model
discussed in Sec.~\ref{sec:H_and_symm}.

\subsection{PBC spectrum}
\label{Appendix:lengthy}

The PBC spectrum of the BdG Hamiltonian in Eq.~\eqref{Eq:H_pbc} is given by

\begin{widetext}
    \begin{align}
E_{\lambda,\epsilon}^{\pm}(k) &= \mp \frac{1}{4} \Bigg\{ 
    -g_{\inter}^2 - g_{\intra}^2 - 2g_2^2 + 16t_{\inter}^2 + 16t_{\intra}^2 + 32t_2^2 - 4\Gamma_+^2 - 4\Gamma_-^2 + 16\Delta_-^2 + 16\Delta_+^2
    + \big(2g_{\inter}g_{\intra} + 32t_{\inter}t_{\intra} + 32it_2\Gamma_+\big)\cos(k a_0) \nn
    &\qquad \qquad + \big(2g_2^2 + 32t_2^2\big)\cos(2k a_0) - \epsilon 8i\big(g_{\intra}t_{\inter} + g_{\inter}t_{\intra}\big)\sin(k a_0)   + \epsilon 8g_2\Gamma_+\sin(k a_0) - \epsilon 16ig_2t_2\sin(2k a_0) \nn
    &\qquad -\lambda 2\sqrt{2} \Bigg[ 
        g_{\inter}^2g_2^2 + g_{\intra}^2g_2^2 - 16g_2^2t_{\inter}^2 - 16g_2^2t_{\intra}^2 - 16g_{\inter}^2t_2^2 - 16g_{\intra}^2t_2^2 + 256t_{\inter}^2t_2^2 + 256t_{\intra}^2t_2^2 - 16ig_{\intra}g_2t_{\inter}\Gamma_+ - 16ig_{\inter}g_2t_{\intra}\Gamma_+ \nn
        &\qquad \qquad \qquad + 16ig_{\inter}g_{\intra}t_2\Gamma_+ + 256it_{\inter}t_{\intra}t_2\Gamma_+ + 2g_{\inter}^2\Gamma_+^2 + 2g_{\intra}^2\Gamma_+^2 - 32t_{\inter}^2\Gamma_+^2 - 32t_{\intra}^2\Gamma_+^2 + 4g_2^2\Gamma_-^2 - 64t_2^2\Gamma_-^2 + 8\Gamma_+^2\Gamma_-^2 \nn
        &\qquad \qquad \qquad - 8g_{\inter}^2\Delta_-^2 - 8g_{\intra}^2\Delta_-^2 + 128t_{\inter}^2\Delta_-^2 + 128t_{\intra}^2\Delta_-^2 - 64\Gamma_+\Gamma_-\Delta_-\Delta_+ + 128\Delta_-^2\Delta_+^2 \nn
        &\qquad \qquad \qquad + \bigg( -16ig_{\inter}^2t_2\Gamma_+ - g_{\inter}\Big\{ 32g_2t_{\intra}t_2 + g_{\intra}\big[ g_2^2 + 4( -12t_2^2 + \Gamma_+^2 - 4\Delta_-^2 ) \big] \Big\}
        - 16\Big\{ g_2^2t_{\inter}t_{\intra} + 2g_{\intra}g_2t_{\inter}t_2
        \nn
        &\qquad \qquad \qquad \qquad \qquad 
        - 16it_{\inter}^2t_2\Gamma_+ - 4t_{\inter}t_{\intra}( 12t_2^2 - \Gamma_+^2 + 4\Delta_-^2 )
        + it_2\big[ g_{\intra}^2\Gamma_+ - 16t_{\intra}^2\Gamma_+ + 4\Gamma_-( \Gamma_+\Gamma_- - 4\Delta_-\Delta_+ ) \big] \Big\} \bigg) \cos(k a_0)   \nn
         &\qquad \qquad \qquad   + \Big[ -g_{\inter}^2( g_2^2 + 16t_2^2 ) - g_{\intra}^2( g_2^2 + 16t_2^2 ) + 16ig_{\intra}g_2t_{\inter}\Gamma_+ + 16ig_{\inter}( g_2t_{\intra} + g_{\intra}t_2 )\Gamma_+ \nn
        &\qquad \qquad \qquad \qquad  + 4g_2^2( 4t_{\inter}^2 + 4t_{\intra}^2 - \Gamma_-^2 ) + 64t_2( 4t_{\inter}^2t_2 + 4t_{\intra}^2t_2 + 4it_{\inter}t_{\intra}\Gamma_+ - t_2\Gamma_-^2 ) \Big] \cos(2k a_0) \nn 
        &\qquad \qquad \qquad  + \big( g_{\inter}g_{\intra}g_2^2 + 16g_2^2t_{\inter}t_{\intra} + 32g_{\intra}g_2t_{\inter}t_2 + 32g_{\inter}g_2t_{\intra}t_2 + 16g_{\inter}g_{\intra}t_2^2 + 256t_{\inter}t_{\intra}t_2^2 \big) \cos(3k a_0) \nn 
        &\qquad \qquad \qquad + \epsilon \Big( 12ig_{\intra}g_2^2t_{\inter} + 12ig_{\inter}g_2^2t_{\intra} - 8ig_{\inter}g_{\intra}g_2t_2 - 128ig_2t_{\inter}t_{\intra}t_2 - 64ig_{\intra}t_{\inter}t_2^2 - 64ig_{\inter}t_{\intra}t_2^2 
        \nn 
                &\qquad \qquad \qquad \qquad  - 4g_{\inter}^2g_2\Gamma_+ - 4g_{\intra}^2g_2\Gamma_+ + 64g_2t_{\inter}^2\Gamma_+ + 64g_2t_{\intra}^2\Gamma_+ + 16ig_{\intra}t_{\inter}\Gamma_+^2 + 16ig_{\inter}t_{\intra}\Gamma_+^2 \nn 
        &\qquad \qquad \qquad \qquad  - 16g_2\Gamma_+\Gamma_-^2 - 64ig_{\intra}t_{\inter}\Delta_-^2 - 64ig_{\inter}t_{\intra}\Delta_-^2 + 64g_2\Gamma_-\Delta_-\Delta_+ \Big) \sin(k a_0)
        \nonumber
\end{align}
\end{widetext}
\begin{widetext}
    \begin{align}  
        &\qquad \qquad \qquad  + \epsilon \Big( 8ig_{\inter}^2g_2t_2 + 8ig_{\intra}^2g_2t_2 - 128ig_2t_{\inter}^2t_2 - 128ig_2t_{\intra}^2t_2 + 4g_{\inter}g_{\intra}g_2\Gamma_+ \nn 
        &\qquad \qquad \qquad \qquad  + 64g_2t_{\inter}t_{\intra}\Gamma_+ + 64g_{\intra}t_{\inter}t_2\Gamma_+ + 64g_{\inter}t_{\intra}t_2\Gamma_+ + 32ig_2t_2\Gamma_-^2 \Big) \sin(2k a_0) \nn
        &\qquad \qquad \qquad   -\epsilon \Big( 4ig_{\intra}g_2^2t_{\inter} + 4ig_{\inter}g_2^2t_{\intra} + 8ig_{\inter}g_{\intra}g_2t_2 
        + 128ig_2t_{\inter}t_{\intra}t_2   + 64ig_{\intra}t_{\inter}t_2^2 + 64ig_{\inter}t_{\intra}t_2^2 \Big) \sin(3k a_0) \Bigg]^{1/2}  \Bigg\}^{1/2},
\label{Eq:PBC_spec_general_case}
\end{align}
\end{widetext}
where the indices $\lambda , \, \varepsilon \in \{+,-\}$ and the overall sign label the eight energy bands. By setting $t_2=g_2=\Delta_-=\Gamma_{-}=0$, Eq.~\eqref{Eq:PBC_spec_general_case} recovers the energy spectrum of the model considered in Ref.~\onlinecite{Chang:2026}.
The complex form in Eq.~\eqref{Eq:PBC_spec_general_case} makes a direct derivation of the gap-closing conditions from the PBC spectrum rather involved and, in general, requires numerical analysis, highlighting the advantages of the coefficient-based approach.

\begin{table}[t]
\centering
\caption{Internal symmetries preserved by Eq.~\eqref{Eq:H_pbc}, together with the symmetry relations and their matrix representations. For notational simplicity, $H$ is used to denote $H_{\rm NHTSC}^{\rm pbc}$ within this table.
Here $H^{*}$ and $H^{T}$ denote complex conjugation and  transpose, respectively. 
In the third column,   
the sign $(\pm)$ denotes whether the square of the corresponding antiunitary symmetry operator is $+1$ or $-1$. For compactness, identity matrices in the Pauli-matrix subspaces ($\mu=0$) are omitted.
}
\begin{tabular}{c c c}
\hline
\hline
Symmetries   &  Relations ~\cite{Kawabata:2019} &  Matrix representations  \\ \hline
PHS & $U_{C_-} H^{T}(k) U_{C_-}^{\dagger}= -H(-k)$ & $U_{C_-} \in \left\{ \eta^{x} (+), \; \eta^{y}  \sigma^{z}(-)\right\}$\\ \hline
TRS$^{\dagger}$ & $U_{C_+} H^{T}(k) U_{C_+}^{\dagger} = H(-k)$&$U_{C_+} \in \left\{ \eta^{z} \sigma^{x}(+), \; \sigma^{y}(-)\right\}$ \\ \hline
SLS &  $U_{S} H(k) U_{S}^{\dagger} = -H(k)$ & $U_{S} \in \left\{ \eta^{y}  \sigma^{x}, \; \eta^{x}  \sigma^{y}\right\}$ \\ \hline
Unitary &  $UH(k)U^{\dagger}= H(k)$ & $U \in \left\{  \; \eta^{z} \sigma^{z}\right\}$   \\ 
\hline \hline
\end{tabular}
\label{Table:Symmetries_without_perturbation}
\end{table}

\subsection{Symmetries and topological properties}
\label{sec:App_symm}

In this section, we discuss the symmetry properties of the 1D NHTSC model
$H_{\rm NHTSC}^{\rm pbc}$ given in Eq.~\eqref{Eq:H_pbc}, both in the absence and in the presence of the perturbation, as well as the onsite disorder term  in Eq.~\eqref{Eq:disorder_term}.
We first consider the unperturbed PBC Hamiltonian. Table~\ref{Table:Symmetries_without_perturbation} summarizes the internal symmetries preserved by $H_{\rm NHTSC}^{\rm pbc}$. Owing to the presence of a nontrivial unitary symmetry, denoted by $U=\eta^z\tau^0\sigma^z$, the Hamiltonian can be block-diagonalized.

After the block diagonalization, as discussed in Sec.~\ref{sec:H_reduction}, the original Hamiltonian is decomposed into two $4\times4$ blocks, $H_\pm(k)$ given in Eq.~\eqref{eq:V_H_V}. They both preserve the SLS, with the unitary operator,
$U_S=\rho_\pm^y\omega_\pm^0$,  
where $\rho_\pm^\mu$ and $\omega_\pm^\mu$ denote Pauli matrices acting on the corresponding subspaces. Accordingly, these blocks belong to class A of the complex AZ with SLS.

In the presence of the perturbation, the nontrivial unitary symmetry is broken, and the PHS, TRS$^\dagger$ and SLS are retained with the following matrix representations:
\begin{subequations}
\label{eq:Symm_pert_H_dhx}
    \begin{eqnarray}
         {\rm PHS}:  U_{C_-} & =& \eta^{x} \tau^{0} \sigma^{0}(+),\\   {\rm TRS}^{\dagger} :  U_{C_+} & =& \eta^{z} \tau^{0} \sigma^{x}(+), \\
    {\rm SLS}:  U_{S} & = &\eta^{y} \tau^{0} \sigma^{x}.
    \end{eqnarray}
\end{subequations}
The symmetries listed in Eq.~\eqref{eq:Symm_pert_H_dhx} classify the  perturbed system into class D within the real AZ classification and class AI$^\dagger$ within the real AZ$^\dagger$ classification, with SLS of $S_{+}$ type.

We now examine the symmetry properties of the onsite disorder term in Eq.~\eqref{Eq:disorder_term}.  In the real-space  BdG representation, the disorder is block-diagonalized inthe form of $\frac{1}{2}\sum_j \Psi^\dagger_j D_j \Psi_j$, where 
\begin{align}
    D_j =\frac{\delta\mu_{+,j}}{2} \eta^z  \tau^0\sigma^0 +  \frac{\delta\mu_{-,j}}{2}\eta^z \sigma^0 \tau^z 
\label{eq:Dj}
\end{align} 
represents the block associated with the unit-cell $j$. In Eq.~\eqref{eq:Dj},  $\delta\mu_{+,j}
\equiv \frac{\delta\mu_{a,j}+\delta\mu_{b,j}}{2}$ and $\delta\mu_{-,j} \equiv \frac{\delta\mu_{a,j}-\delta\mu_{b,j}}{2}$ are the symmetric and asymmetric components of the onsite disorder potential terms, respectively.
Since $\delta\mu_{a,j}$ and $\delta\mu_{b,j}$ are real, the onsite disorder matrix satisfies $D_j^T=D_j$. One can show that the disorder term  preserves the relevant internal symmetries. Specifically, using the real-space form~\cite{Kawabata:2019} of the symmetry operators in Eq.~\eqref{eq:Symm_pert_H_dhx}, we find
\begin{subequations}
    \begin{align}
       \text{PHS:}\quad &U_{C_-} D_j^T U_{C_-}^{\dagger} = -D_j,\\
       \text{TRS}^{\dagger}\text{:}\quad &U_{C_+} D_j^T U_{C_+}^{\dagger} = D_j,\\
       \text{SLS:}\quad &U_S D_j U_S^{\dagger} = -D_j.
    \end{align}
\end{subequations} 
Therefore, although the onsite disorder breaks the translation symmetry, it preserves the internal symmetries that protect the zero modes. Consequently, the zero modes can persist as long as the bulk gap does not close, consistent with the numerical results presented in the main text.

\begin{figure*}[t]
    \centering
    \includegraphics[width=0.95\textwidth]{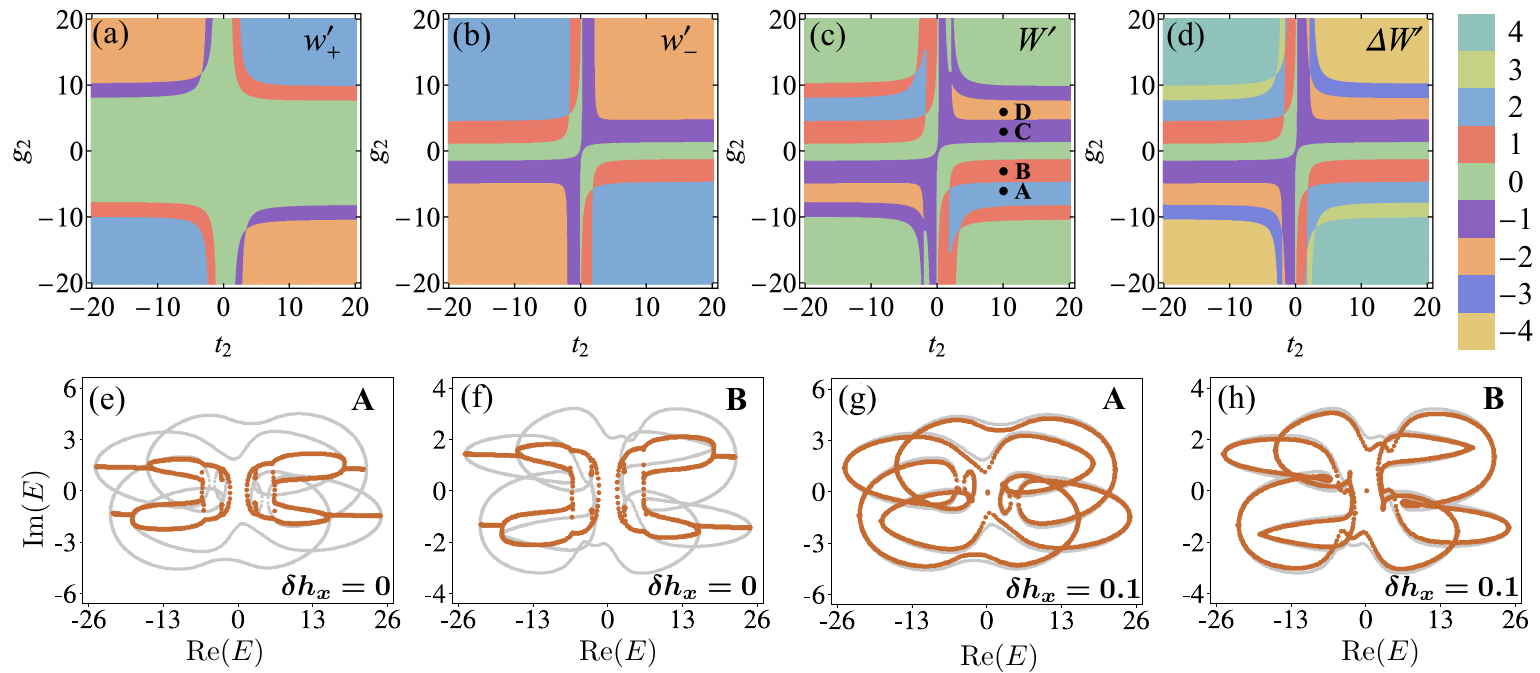}
     \caption{(a--d) Phase diagrams characterized by the subsystem and composite winding numbers, $w_+^\prime$, $w_-^\prime$,  $W^\prime \equiv w_+^\prime +w_-^\prime$, and $ \Delta W^\prime \equiv \Wnhse_- (0) \equiv w_-^\prime - w_+^\prime$, respectively.
     (e)--(h) Energy spectra obtained with $N=500$ and for  (e,f) $\delta h_x = 0$  and (g,h)  $\delta h_x = 0.1$. Points A--D in panel (c) mark the same parameter sets $(t_2, \, g_2)$ as those indicated in Fig.~\ref{fig:phase_diagram}. Panels (e,g) and (f,h) correspond to parameter sets A and B, respectively. The adopted values of the remaining parameters are given in the caption of Fig.~\ref{fig:phase_diagram}.
    }
    \label{fig:Supp_B03_PD}
\end{figure*}

\begin{figure}[ht]
    \centering
    \includegraphics[width=0.5\textwidth]{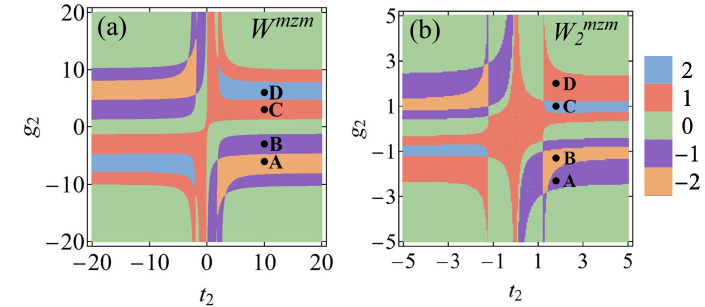}
    \caption{Phase diagrams characterized by (a) the perturbed winding number $W^{\rm mzm}$ for the model with sublattices and (b) the perturbed winding number $W^{\rm mzm}_2$ for the model without sublattices, analogous to $W^{\rm mzm}$ of Eq.~\eqref{eq:W_by_V}. The remaining parameters in panels (a) and (b) are the same as those used in Figs.~\ref{fig:phase_diagram} and~\ref{Fig:NHTSC2_phase_diagram}, respectively. The perturbation used in this figure is $\delta h_x =0.05$.} 
    \label{fig:PD_dhx}
\end{figure}

\begin{figure}[ht]
    \centering
    \includegraphics[width=0.48\textwidth]{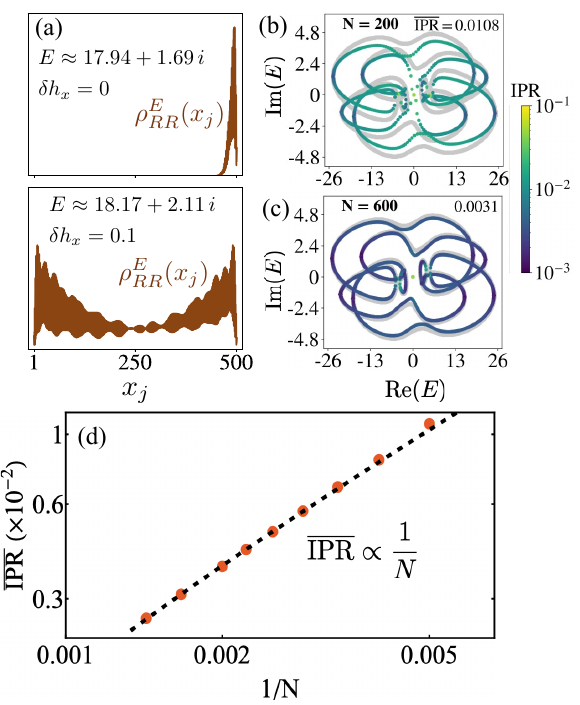}
    \caption{(a) Total spatial density profiles of bulk states at the indicated, nearly equal energies $E$. The upper and lower panels correspond to $\delta h_x=0$ and $\delta h_x=0.1$, respectively, with $N=500$.  (b,c) OBC and PBC spectra of the model described by Eq.~\eqref{Eq:H_NHTSC}, with (b) $N=200$ and (c) $N=600$. (d) Average IPR computed from Eq.~\eqref{eq:IPR_avg} as a function of $1/N$. The dashed line is a fit to $\overline{\rm IPR}\propto 1/N$. All panels are obtained using the same parameters as those specified in the caption of Fig.~\ref{fig:phase_diagram}(c).
    } 
    \label{fig:OBC_varyN_avgIPR}
\end{figure}

\subsection{Analytical expression for phase boundaries}

We employ the approach described in Appendix~\ref{Appendix:deducing_phase_boundaries} to derive the analytical expressions for the topological phase boundaries from $H_{\rm NHTSC}^{\rm pbc}(k)$. Specifically, by applying Eq.~\eqref{Eq:resultant_phase_boundary}, we obtain the conditions for the phase boundaries directly from the $2\times2$ blocks $h_{\pm}$ introduced in Sec.~\ref{sec:H_reduction}, without solving the polynomial equation explicitly. The final results are algebraic equations written in terms of the model parameters. Because of their lengthy form, we provide them in a separate file available in Ref.~\onlinecite{PhaseboundaryFile}. We have checked that these expressions give results consistent with the winding-number calculations throughout this work.
 
In addition, we separately derive the phase-boundary conditions associated with the special point $z=-1$, corresponding to $k a_0=\pm \pi$. We confirm both analytically and numerically that these $z=-1$ conditions are included in the full phase-boundary expressions in Ref.~\onlinecite{PhaseboundaryFile} and do not generate additional independent phase boundaries.

\section{Additional numerical results}
\label{sec:additional_results}

In this section, we present additional numerical results.
As a consistency check supplementing Fig.~\ref{fig:phase_diagram} in the main text, we compute $w^\prime_\pm$ [Eq.~\eqref{eq:w_pm_prime}], $W^\prime \equiv w^\prime_+ + w^\prime_- $ and $\Wnhse_- (0) = w_- + w_-^\prime = w_- - w_+ = - \Delta W $ [Eq.~\eqref{Eq:W_nhse}] and present the results in  Fig.~\ref{fig:Supp_B03_PD}. 
We also include the energy spectra for the parameter sets corresponding to A and B labeled in Fig.~\ref{fig:phase_diagram}. 
In addition, to confirm the prediction in Sec.~\ref{sec:topo-inv} of the main text that the perturbed winding number recovers its unperturbed value, we compute the perturbed winding numbers via Eq.~\eqref{eq:W_by_V} for the NHTSC models with and without sublattice. The corresponding phase diagrams are shown in Figs.~\ref{fig:PD_dhx}(a) and (b), respectively. With all other parameters fixed, the overall structures of the perturbed phase diagrams are almost unchanged compared with the unperturbed ones in Figs.~\ref{fig:phase_diagram} and~\ref{Fig:NHTSC2_phase_diagram}, respectively, apart from small shifts in the phase boundaries. This confirms that the weak perturbation suppresses the NHSE without changing the underlying topological properties relevant to the MZMs.

To verify the suppression of the NHSE, we show in Fig.~\ref{fig:OBC_varyN_avgIPR}(a) the total spatial density profiles of representative bulk-state wave functions with and without the perturbation field. For a fair comparison, the two states are chosen at nearly the same energy, since the spectra with and without the perturbation are generally not identical. When the perturbation field is introduced, the bulk-state density profile changes from a boundary-localized distribution to a spatially extended one. We further examine the system-size dependence of the OBC spectra in the presence of the perturbation field. As shown in Figs.~\ref{fig:OBC_varyN_avgIPR}(b) and (c), the bulk OBC spectra progressively approach the corresponding PBC spectrum as the system size increases, suggesting the suppression of the NHSE. This conclusion is further supported by the average IPR, defined in Eq.~\eqref{eq:IPR_avg}, over the entire OBC spectrum in Fig.~\ref{fig:OBC_varyN_avgIPR}(d), which follows $\overline{\rm IPR}\propto 1/N$~\cite{Mirlin:RMP2008,Park:2016PRB}, consistent with extended bulk states dominating the OBC spectrum.

\begin{figure*}[h]
    \centering
    \includegraphics[width=0.95\textwidth]{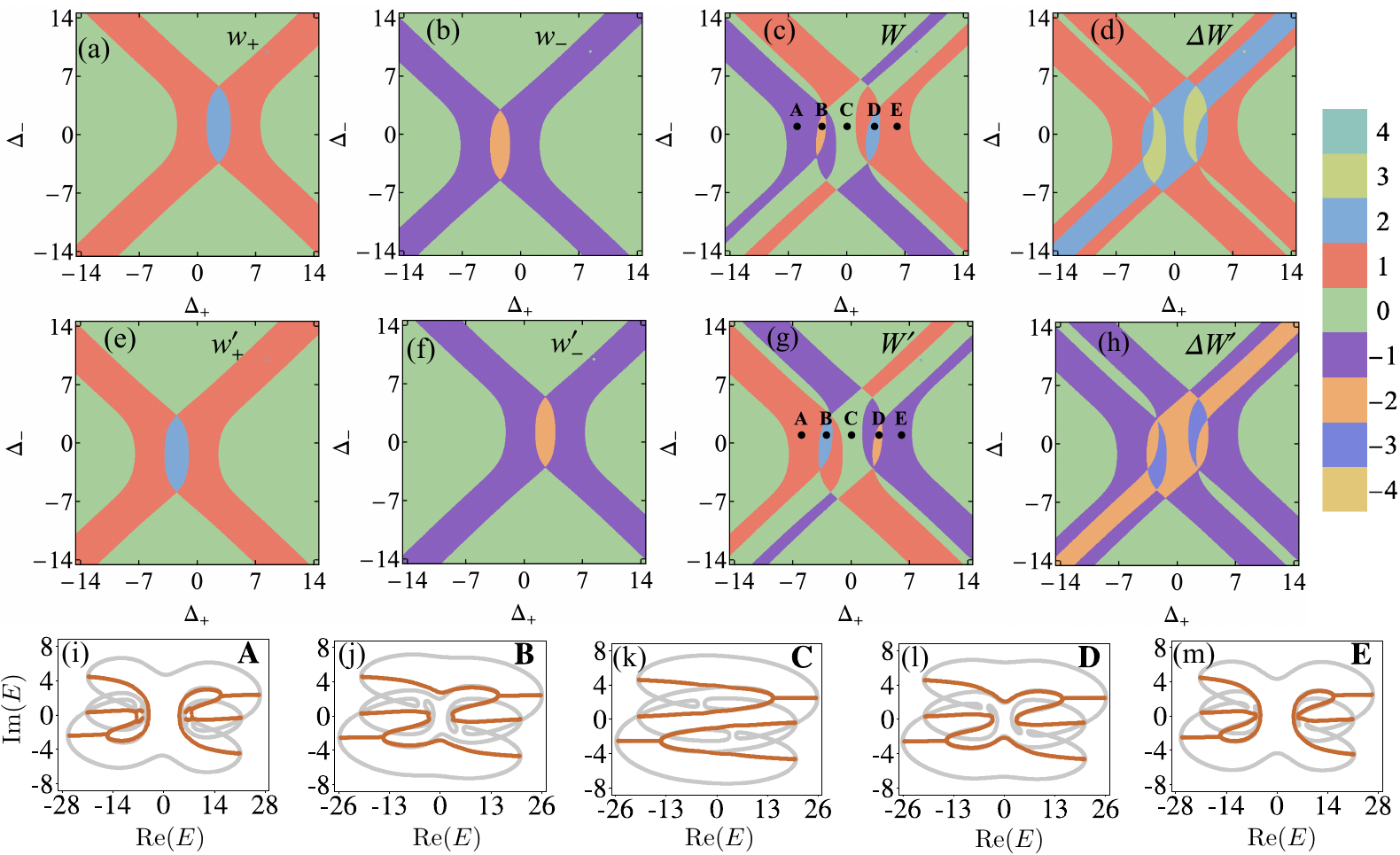}
    \caption{(a)--(h) Phase diagrams in the ($\Delta_+,\,  \Delta_-$) plane, analogous to Figs.~\ref{fig:phase_diagram} and \ref{fig:Supp_B03_PD}(a--d), for $4t_+ = 12$, $4t_- = 4.0$, $g_+ = 6.0$, $g_- = 3.0$, $t_2 = 10.0$, $g_2 = 6.0$, $\Gamma_+ = 5.0$, and $\Gamma_- = 2.5$. Points A--E denote ($\Delta_+, \, \Delta_-$) with fixed $\Delta_- = 1.0$ and $\Delta_+ = \{-6,\,-3,\,0,\,3.3,\,6.0\}$. (i)--(m) PBC (gray) and OBC (brown) energy spectra for the parameter sets A--E with $\delta h_x = 0$ and $N=500$. 
    }
    \label{fig:Supp_E03_PD}
\end{figure*}

\begin{figure*}[t]
    \centering
    \includegraphics[width=0.95\textwidth]{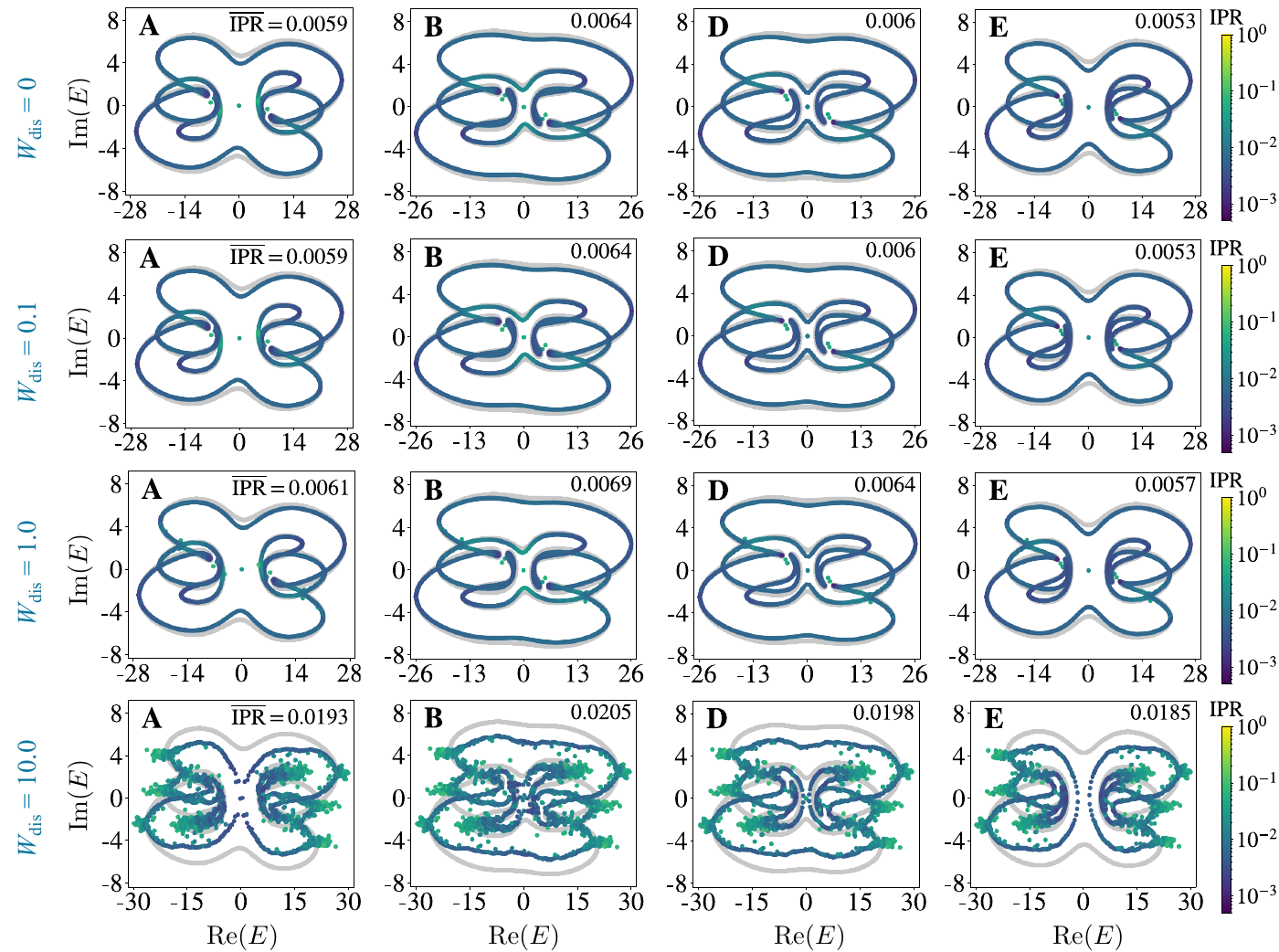}
    \caption{Similar plots to Fig.~\ref{fig:spectra_Dis}, but with different  parameter sets; points A, B, D, and E correspond to the parameter sets indicated in Fig.~\ref{fig:Supp_E03_PD}(c). The OBC spectra are obtained using a single disorder realization.  
    }
    \label{fig:spectra_Dis_E03}
\end{figure*}

In addition to the phase diagrams in the $(t_2,g_2)$ plane in the main text, we explore phase diagrams in the $(\Delta_+,\Delta_-)$ plane, as presented in 
Figs.~\ref{fig:Supp_E03_PD}(a)--(h).
The energy spectra in the clean limit corresponding to the parameter sets A to E are displayed in Figs.~\ref{fig:Supp_E03_PD}(i)--(m), while those  in the presence of disorder are presented in Fig.~\ref{fig:spectra_Dis_E03}, along with the IPR and the average \(\overline{\mathrm{IPR}}\).

Consistent with the main text, the appearance of the MZMs, including the phases with multiple MZM pairs, is reflected in the winding numbers computed from the Schur-Cohn method. Furthermore, the MZMs are robust against weak disorder on the scale of the bulk gap.  

\clearpage


%

\end{document}